\documentclass[aps, 10pt,
notitlepage, twocolumn, superscriptaddress,
nofootinbib]{revtex4-1}

\usepackage{amsmath}
\usepackage{amssymb}
\usepackage{amsfonts}
\usepackage[utf8]{inputenc}
\usepackage[T1]{fontenc}
\usepackage{mathrsfs}
\usepackage{anyfontsize}

\usepackage[linktocpage,breaklinks]{hyperref}
\usepackage[usenames,dvipsnames]{xcolor}

\usepackage{graphicx}
\usepackage{epsfig}
\usepackage{epstopdf}
\usepackage{natbib}
\usepackage{hyperref}
\usepackage{orcidlink}

\hypersetup{colorlinks=true, citecolor=Purple, linkcolor=Purple,
urlcolor=Purple}

\newcommand{\GB}{\mathscr{G}}
\newcommand{\rh}{r_h}

\usepackage{tikz}

\begin{document}

\allowdisplaybreaks

\title{Axial perturbations of hairy black holes in generalised scalar-tensor theories}

\author{{Georgios~Antoniou}}
\affiliation{Dipartimento di Fisica, ``Sapienza'' Universit\'a di Roma \& Sezione INFN Roma1, P.A. Moro 5, 00185, Roma, Italy}
\affiliation{Dipartimento di Fisica,  Universit\'a di Pisa, 56127 Pisa, Italy}

\author{{Caio F. B. Macedo}}
\affiliation{Faculdade de Física, Universidade Federal do Pará, Salinópolis, Pará, 68721-000 Brazil}

\author{{Andrea~Maselli}}
\affiliation{Gran Sasso Science Institute (GSSI), I-67100 L’Aquila, Italy}

\author{{Thomas~P.~Sotiriou}}
\affiliation{Nottingham Centre of Gravity \&
School of Mathematical Sciences, University of Nottingham,
University Park, Nottingham NG7 2RD, United Kingdom}
\affiliation{School of Physics and Astronomy, University of Nottingham,
University Park, Nottingham NG7 2RD, United Kingdom}

\begin{abstract}
    Gravitational wave observations can test the validity of General Relativity (GR) in the strong field regime. Certain classes of scalar-tensor theories indeed predict that compact objects can exhibit significant deviations from their GR counterparts. Here we explore the quasinormal modes of axial perturbations in spherically symmetric black holes in three such classes: (i) dilatonic black holes with an additional scalar-Ricci coupling (EdRGB), (ii) spontaneously scalarized black holes (EsRGB) with a quadratic coupling to the Gauss-Bonnet invariant and the Ricci scalar, (iii) spontaneously scalarized black holes with a quadratic and a quartic coupling to the Gauss-Bonnet invariant.
\end{abstract}

\maketitle


%
\section{Introduction}
%

\label{sec:introduction}
The first detection of gravitational waves \cite{LIGOScientific:2016aoc} marked the beginning of gravitational-wave-astronomy.
We already have an impressive number of observations at hand \cite{LIGOScientific:2016aoc,LIGOScientific:2018mvr,LIGOScientific:2020ibl,LIGOScientific:2021usb,LIGOScientific:2021djp}, which is expected to grow significantly by the time the 4th observational run has concluded, with KAGRA now joining the LIGO and VIRGO collaborations. This unprecedented access to the strong-field regime of gravitation allows us to confront potential deviations from GR manifesting in these high curvature regions, and to test the existence of additional fields.

A scalar field nonminimally coupled to gravity would be perhaps the simplest example of a new field that leads to a modification of GR.
In recent years, various black hole solutions have been derived in the context of scalar-Gauss-Bonnet gravity, where a scalar couples to the Gauss-Bonnet curvature invariant. These theories constitute sub-classes of more general scalar-tensor theories belonging in the Horndeski framework \cite{Horndeski:1974wa,Deffayet:2009mn,Kobayashi:2011nu}.
Scalar-Gauss-Bonnet gravity exhibits significant deviations from GR in the strong gravitational field regimes, while passing high accuracy tests of GR in the weak-field approximation \cite{Elder:2022rak}.
In particular, black holes in scalar-Gauss-Bonnet gravity evade no-hair theorems \cite{Bekenstein:1995un,Hawking:1972qk,Sotiriou:2011dz,Silva:2017uqg} and can carry scalar-hair \cite{Mignemi:1992pm,Kanti:1995vq,Sotiriou:2013qea,Antoniou:2017acq,Silva:2017uqg,Doneva:2017bvd,Antoniou:2017acq}, which is sourced by the nontrivial interactions of the scalar field with the Gauss-Bonnet invariant.

Different types of couplings between the scalar and the Gauss-Bonnet invariant allow for black holes with different characteristics. For linear coupling all black holes will have scalar hair. This coupling is known to be the only shift-symmetric interaction that can lead to black hole hair \cite{Sotiriou:2013qea,Sotiriou:2014pfa}, which is particularly relevant for massless (or almost massless) fields. However, it can also be thought of as a the weak field or weak coupling limit of the exponential coupling, which arises as a limit of heterotic string theory, and has received a lot of attention in the context of black holes \cite{Mignemi:1992pm,Kanti:1995vq,Yunes:2011we,Maselli:2015tta}. A quadratic coupling, or more general couplings that allow for GR solutions, can instead lead to black hole scalarization: whether black holes will have hair or not depends on their mass \cite{Silva:2017uqg,Doneva:2017bvd} or spin \cite{Dima:2020yac}. 

An important feature of models that exhibit black hole scalarization, is that only a few interaction terms that appear in linear perturbation theory around GR determine whether scalarization occurs \cite{Andreou:2019ikc}, whereas the properties of scalarized black hole solution are very sensitive to the specific nonlinear interaction included in the model \cite{Blazquez-Salcedo:2018jnn,Silva:2018qhn,Macedo:2019sem,Antoniou:2021zoy,Antoniou:2022agj}.
Hence, the viability of such models depends crucially on the type of nonlinear interactions that are present. There has been a considerable amount of work on all aspects of theories that exhibit scalarization and of scalarized black holes, including static and spinning solutions \cite{Mignemi:1992pm,Kanti:1995vq,Silva:2017uqg,Antoniou:2022agj,Ayzenberg:2014aka,Herdeiro:2020wei}, their stability \cite{Kanti:1997br,Blazquez-Salcedo:2018jnn,Silva:2018qhn,Macedo:2019sem,Macedo:2020tbm,Blazquez-Salcedo:2020rhf,Minamitsuji:2024twp}, observational imprints \cite{Wong:2022wni}, as well as determining whether they can be simulated in the context of a well-posed initial value problem with numerical relativity techniques \cite{Ripley:2019aqj,East:2020hgw,East:2021bqk}. See also Ref.~\cite{Doneva:2022ewd} for a review.

One particular interaction, namely a coupling between the scalar field and the Ricci curvature, has been shown to improve the behaviour and properties of scalarization models and their black holes on multiple counts. It stabilizes otherwise unstable scalarized black hole solutions \cite{Antoniou:2022agj}; it prevents neutron stars from scalarizing \cite{Ventagli:2021ubn}, thereby removing pulsar constraints; it allows for a late-time GR cosmological attractor \cite{Antoniou:2020nax}, thereby making scalarization cosmologically viable; it improves the hyperbolicity of the linear perturbation equations \cite{Antoniou:2022agj} and mitigates the loss of hyperbolicity in spherical collapse \cite{Thaalba:2023fmq}. It is therefore particularly interesting to study quasi-normal modes (QNMs) for black holes in theories that contain this interaction.

The QNM spectrum of dilatonic black holes has been studied in \cite{Blazquez-Salcedo:2017txk,Blazquez-Salcedo:2016enn} 
and in \cite{Pierini:2022eim} for static and slowly rotating 
solutions, respectively. In such works 
the Einstein-Hilbert action is supplemented by the scalar-GB interaction and a canonical kinetic term for the scalar. Polar and axial perturbations have also been studied in the context of spontaneous scalarization with a quadratic exponential coupling in \cite{Blazquez-Salcedo:2020rhf,Blazquez-Salcedo:2020caw}. Here we will consider an action that contains couplings of the scalar to both the Gauss-Bonnet invariant and the Ricci scalar. This generalizes earlier work and will allow us to analyse the effects of the Ricci coupling to the QNMs for the first time.

The paper is structured as follows: in Sec.~\ref{sec:theory} we introduce the theoretical framework of our analysis, containing the dilatonic and spontaneously scalarized black holes as special cases. In Sec.~\ref{sec:bg_and_pert} we discuss the numerical background solutions emerging in this framework, and we introduce linear order axial perturbations. We present the axial potential and discuss two different methods of searching for the QNMs, namely integration in the time and frequency domains. In Sec.~\ref{sec:results} we present the main results of our analysis. Finally, in Sec.~\ref{sec:conclusions} we present our conclusions.

%
\section{Theoretical setup}
\label{sec:theory}
%
We consider the following theory
\begin{equation}
\label{eq:Action}
    S=\frac{1}{2\kappa}\int d^4 x \sqrt{-g}\;\bigg[ R + X 
    + J(\phi)R + F(\phi)\GB\bigg]\, ,
\end{equation}
where $\kappa=8\pi G$, $X=-(\nabla\phi)^2/2$ is the scalar-field kinetic term, and $\GB \equiv R_{\mu\nu\rho\sigma} R^{\mu\nu\rho\sigma} -4R_{\mu\nu} R^{\mu\nu} + R^2$ is the Gauss-Bonnet 
invariant. 
The field equations for the metric and for the scalar 
field obtained by varying the action are given by:
\begin{equation}
\begin{split}
    &\big[1+J(\phi)\big]G_{\mu\nu}+\frac{1}{4}g_{\mu\nu}(\nabla\phi)^2
    -\frac{1}{2}\nabla_\mu\phi\nabla_\nu\phi
    \\
    &+\left(g_{\mu\nu} \nabla^2 -\nabla_\mu\nabla_\nu \right)J(\phi)
    \\
    &+\frac{1}{g}g_{\mu(\rho}g_{\sigma)\nu}\epsilon^{\kappa\rho\alpha\beta}\epsilon^{\sigma\gamma\lambda\tau}R_{\lambda\tau\alpha\beta}\nabla_{\gamma}\nabla_{\kappa}F(\phi)=0\, ,
\end{split}
\label{eq:bg_grav}
\end{equation}
and
\begin{equation}
\nabla_\mu \nabla^\mu \phi +(\partial_{\phi} J )R +(\partial_{\phi} F)\GB=0\, ,
\label{eq:bg_scalar}
\end{equation}
where $\partial_\phi$ denotes derivatives with respect to $\phi$.

Although the action we are considering is not a complete effective field theory (EFT), it arguably contains the most interesting terms for strong field phenomenology in such an EFT. As already discussed in the introduction, the coupling between the scalar and the Gauss-Bonnet invariant leads to hairy black holes \cite{Kanti:1995vq,Antoniou:2017acq,Doneva:2017bvd,Silva:2017uqg,Antoniou:2017hxj,Bakopoulos:2018nui,Ramazanoglu:2016kul,Blazquez-Salcedo:2018jnn,Silva:2018qhn,Macedo:2019sem,Herdeiro:2018wub,Ramazanoglu:2017xbl,Ramazanoglu:2018hwk,Dima:2020yac,Herdeiro:2020wei,Berti:2020kgk,Antoniou:2021zoy,Thaalba:2022bnt}.
The coupling with the Ricci scalar does not generate hairy solutions but it modifies the properties of the ones obtained from the Gauss-Bonnet coupling \cite{Antoniou:2020nax,Ventagli:2021ubn,Antoniou:2022agj,Thaalba:2023fmq}. These two couplings could also be considered responsible for breaking shift symmetry for the scalar in curved spacetimes.

Action~\eqref{eq:Action} covers a very large theory space so in practice we focus on specific interesting sections of it:
\begin{enumerate}
    \item \textbf{Einstein-dilaton-Ricci-Gauss-Bonnet (EdRGB)}. This case corresponds to the coupling functions $J(\phi)=-\beta\,\phi^2/4\quad\text{and}\quad F(\phi)=\alpha\,\exp(\phi)/4$. The BHs belonging to this family have secondary hair. In absence of the coupling with the Ricci term, i.e. $\beta\to 0$, the theory reduces to that of EdGB dilatonic black holes which are often viewed as a low-energy limit of some string theory \cite{Kanti:1995vq}.
    \item \textbf{Einstein-scalar-Ricci-Gauss-Bonnet (EsRGB)}. This case corresponds to choosing coupling functions $J(\phi)=\beta\,\phi^2/4\quad\text{and}\quad F(\phi)=\alpha\,\phi^2/2$. This choice leads to scalarized black hole solutions. Their properties and radial stability have been studied 
    in \cite{Antoniou:2021zoy,Antoniou:2022agj}, where it was shown that the Ricci term can stabilize the otherwise unstable solutions when $\beta>0$.
    \item \textbf{Einstein-scalar-Gauss-Bonnet (EsGB)}: This case corresponds to the choice $J(\phi)=0\quad\text{and}\quad F(\phi)=\alpha\,\phi^2/2+\zeta\,\phi^4/4$. We consider it to explore the minimum requirements to generate stable spontaneously scalarized BHs with only the Gauss-Bonnet term.  While the quadratic term is enough to spontaneously scalarize BHs, these scalarized black holes would be radially unstable \cite{Blazquez-Salcedo:2018jnn}, and the quartic term in the coupling can render them stable \cite{Silva:2018qhn}. 
\end{enumerate}
The key difference between the first and the other two cases we consider is that in EdRGB gravity all black holes will carry scalar hair, whereas
EsRGB and EsGB gravity exhibit spontaneous scalarization and only certain black holes will have hair. We can describe the main idea behind spontaneous 
scalarization as follows. Let us consider a BH solution in GR, 
and compute linear radial perturbation of the scalar field around 
its constant background value $\phi_0$, \textit{i.e.} $\phi=\phi_0(r)+\delta\phi(t,r,\theta,\varphi)$. Then, from the field equation 
\eqref{eq:bg_scalar} we have 
\begin{equation}
    \Box\delta\phi=-\big[\partial_{\phi}^2 J\,R+\partial_{\phi}^2 F\,\GB\big]_{\phi=\phi_0}\delta\phi\, ,
\end{equation}
where the quantity in square brackets acts as 
an effective mass (squared) 
$m^2_\textnormal{eff}$ for $\phi$.
In this framework GR solutions are possible provided that 
$\phi_0$ exists and that $\partial_{\phi}F|_{\phi=\phi_0}=0$, 
with $\partial_{\phi}^2 F|_{\phi=\phi_0}\GB<0$.
However, if the effective mass becomes sufficiently 
negative (after appropriately fixing the sign of 
the coupling function), it can trigger a tachyonic 
instability leading to spontaneously 
scalarised BH solutions, which branch out from the 
GR limit.

By studying the linear stability of scalar perturbations 
around the GR background, we can find a threshold value for the constant $\alpha=\alpha_{\rm t}$.
We may decompose the scalar perturbations in spherical harmonics $\delta\phi\sim e^{-i\omega t}\phi_1(r)\;Y^m_\ell(\cos\theta)\,e^{-im\varphi}/r$. This allows us to look for instabilities which will grow and eventually lead to the development of scalar hair. These unstable states correspond to solutions of the scalar perturbation equation. for which $d\phi_1/dr|_{r\rightarrow\infty}=0$. The perturbation equation on a Schwarzschild background reads
\begin{equation}
    r(r-\rh)\, \phi_1''+\rh\, \phi_1'
    +\bigg[ \frac{\omega^2r^3}{r-\rh} + \frac{12\alpha \rh^2}{r^4} - \frac{\rh}{r} -\ell(\ell+1) \bigg]\, \phi_1 = 0,
\end{equation}
here $r_h=2M$ is the event horizon for the Schwarzschild spacetime. Such threshold, expressed in terms of the dimensionless ratio 
$\alpha_t/M^2$, depends on the multipolar distribution of the scalar perturbation $\delta \phi$, symbolized by the spherical harmonic index $\ell$, as well as the number of nodes of the scalar field $n$. In Table~~\ref{table:scalarization_thresholds} we show the threshold for given values of $\ell$ and $n$. Note that the smaller threshold $\alpha_t$ for which scalarization occurs is for $(\ell,n)=(0,0)$, and, incidentally, this leads to the faster unstable mode for $\alpha>\alpha_{\rm t}$ in some theories~\cite{Blazquez-Salcedo:2018jnn}.

\begin{table}
    \centering
    \label{table:scalarization_thresholds}
    \begin{tabular}{ c c c c }
    \hline
    \multicolumn{4}{c}{Threshold \hspace{2mm} $\alpha_{\rm t}/M^2$} \\[1mm]
    \hline\hline
    & $\ell=0$\hspace{3mm} & $\ell=1$\hspace{3mm} & $\ell=2$\hspace{3mm} \\[1mm]
    \hline
    $n=0$\hspace{3mm} & 0.719\hspace{3mm} & 2.088\hspace{3mm} & 4.098 \\ $n=1$\hspace{3mm} & 4.960\hspace{3mm} & 7.589\hspace{3mm} & 10.96 \\
    $n=2$\hspace{3mm} & 13.03\hspace{3mm} & 16.94\hspace{3mm} & 21.63\\
    \hline
    \end{tabular}
\caption{Scalarization threshold for various pairs $(n,\ell)$.}
\end{table}

%
\section{Background solutions and linear 
order perturbations}
\label{sec:bg_and_pert}
%

In this work we focus on spherically symmetric 
BH spacetimes
\begin{equation}
    ds^2=-A(r)dt^2+\frac{dr^2}{B(r)}+r^2d\Omega^2,
\label{eq:metric}
\end{equation}
where $d\Omega^2=d\theta^2+\sin^2\theta d\varphi^2$ is the line element of the 2-sphere. In   analytical symbolic computations, it is convenient to use $\chi=\cos\theta$ instead of the $\theta$ to speed up the analysis.

To compute QNMs we need to determine 
the numerical BH background in which the tensor and scalar 
perturbations propagate. The procedure was already thoroughly (see, e.g., Refs.~\cite{Silva:2018qhn,Macedo:2019sem,Antoniou:2022agj}) explored in the literature, but we outline it here for completeness. The background field equations are given in Appendix~\ref{ap:background}.
We start by demanding that the metric components in Eq.~\eqref{eq:metric} have the proper behavior describing a BH. Near the event horizon, this is equivalent to considering the following expansions
\begin{align}
    A(r)=&\sum_{n=1}a^{(n)}(r-r_h)^n, \label{eq:expA}\\
    B(r)=&\sum_{n=1}b^{(n)}(r-r_h)^n , \\
    \phi_0(r)=&\sum_{n=0}\phi^{(n)}(r-r_h)^n ,\label{eq:phi},
\end{align}
where the coefficients $\{a^{(n)},b^{(n)},c^{(n)}\}$ are obtained by inserting the above expansion into the field equations~\eqref{eq:bg_grav}-\eqref{eq:bg_scalar}, expanding close to the horizon, and solving order by order iteratively. Far from the BH, i.e. $r\gg r_h$, we have that
\begin{align}
    A(r)=&1-\frac{2M}{r}+\sum_{n=1}\frac{\tilde{a}^{(n)}}{r^n}\label{eq:expAi}\\
    B(r)=&1-\frac{2M}{r}+\sum_{n=1}\frac{\tilde{b}^{(n)}}{r^n} , \\
    \phi_0(r)=&\frac{Q}{r}+\sum_{n=2}\frac{\tilde{\phi}^{(n)}}{r^n},\label{eq:phii}
\end{align}
where $M$ and $Q$ are the ADM mass and the scalar charge 
respectively. Equations \eqref{eq:expAi}-\eqref{eq:phii} 
are consistent with the requirement that the BH solutions 
are asymptotically flat. Once again,
the coefficients $\{\tilde{a}^{(n)},\tilde{b}^{(n)},\tilde{\phi}^{(n)}\}$ 
can be obtained by replacing $A(r), B(r)$ and $\phi(r)$ into 
Eqs.~\eqref{eq:bg_grav}-\eqref{eq:bg_scalar} and solving 
order by order in $1/r$.

For the theory described by the Lagrangian \eqref{eq:Action}, 
BH solutions that are regular at the horizon are possible provided that the 
following existence condition is satisfied:
\begin{equation}
\begin{split}
    \lim_{r\rightarrow r_h}&
    \big\{
    J^2 r^6-48 (\partial_{\phi} F)^2 r^2 \left[9 (\partial_{\phi} J)^2+2\right]\\
    &
    +2 J [r^6 (3 (\partial_{\phi} J)^2+1)-48 (\partial_{\phi} F)^2 r_h^2]\\
    &
    -1152 (\partial_{\phi} F)^3 \partial_{\phi} J+r^6 [3 (\partial_{\phi} J)^2+1]^2\big\}\ge 0\, ,
\end{split}
\end{equation}
Note that the first derivative of the scalar field 
at the horizon is given in terms of the parameters 
$\alpha,\, \beta$, and of the value of $\phi$ at $r_h$, 
$\phi_h$.

We can now introduce linear order perturbations as
\begin{equation}
g_{\mu\nu}=g_{\mu\nu}^{\text{BG}}+\varepsilon\, \delta h_{\mu\nu}\quad\ ,\quad \phi=\phi_0+\varepsilon\,\delta\phi\,,
\end{equation}
where $\varepsilon \ll 1$ is a bookkeeping parameter representing the order of the perturbation. The quantities with superscript $\text{BG}$ 
correspond to the numerical background solutions and depend only on the radial coordinate $r$, while the perturbations are functions of $(t,r,\chi,\varphi)$. 
The symmetry of the background allows to expand metric 
perturbations into tensor spherical harmonics 
$T^{\ell m}_{\mu\nu}(\chi,\varphi)$, which depend on the 
usual spherical harmonics $Y^{m}_{\ell}$ and their 
derivatives \cite{Regge:1957td,Zerilli:1970wzz}. This 
leads $\delta h_{\mu\nu}$ to separate into axial and 
polar sectors according to their properties under a parity 
transformation. For a spherical background such components 
decouple and can be treated independently
\begin{align}
    \delta h_{\mu\nu}(t,r,\chi,\varphi)=\sum_{\ell m}\big\{ [h^{\ell m}_{\mu\nu}(r,t) T_{\mu\nu}^{\ell m}(\chi,\varphi)]_{\rm ax}+\nonumber\\
    \sum_{\ell m}[h^{\ell m}_{\mu\nu}(r,t) T_{\mu\nu}^{\ell m}(\chi,\varphi)]_{\rm pol}\big\}\ ,
\label{eq:metricpert}
\end{align}
In the same spirit, the scalar perturbation can be 
decomposed in terms the spherical harmonics and reads
\begin{equation}
    \delta\phi=\sum_{\ell m} \frac{\phi^{\ell m}_1(t,r)}{r}Y_{\ell}^{m}(\chi,\varphi) \ .
\end{equation}
Because of the parity properties of the scalar field perturbations the axial metric perturbations decouple from the scalar field perturbations. For the polar sector, however, metric and scalar perturbations couple with each other, similarly to what happens for EdGB BHs~\cite{Blazquez-Salcedo:2016enn}.

In this work we focus on  the axial sector only. Then, adopting the Regge-Wheeler gauge, the components of the metric perturbations in 
Eq.~\eqref{eq:metricpert} take the following form: 
\begin{equation}
    [h_{\mu\nu}^{\ell m}T^{\ell m}_{\mu\nu}]_{\rm ax}=
    \begin{pmatrix}
    0 & 0 & 0 & (\chi^2-1)\,h^{\ell m}_0\\
    0 & 0 & 0 & (\chi^2-1)\, h^{\ell m}_1\\
    0 & 0 & 0 & 0\\
    * & * & 0 & 0
    \end{pmatrix}
    \partial_\chi Y^{lm} \, .
\label{eq:axial_RW_gauge}
\end{equation}

The field equations relevant for computing 
$h^{\ell m}_0,h^{\ell m}_1$  are 
shown in Appendix~\ref{ap:perturbation_equations}.
For the sake of clarity, hereafter we take the sum on 
$(\ell,m)$ to be implicit and we also drop 
the multipolar indices.

%
\subsection{Time and frequency domain integration}
%
By expanding the field equations~\eqref{eq:bg_grav}-\eqref{eq:bg_scalar} 
up to first order in $\varepsilon$, we obtain the 
differential equations for the metric components $h_0$ and $h_1$. We solve such equations in the time and 
frequency domain, resorting to different 
techniques. 

For the time evolution it is useful to cast the 
equations for axial perturbations in terms of a 
single master equation for $h_1$:
\begin{equation}
   g(r)^2\frac{\partial^2 h_1}{\partial t^2} -\frac{\partial^2 h_1}{\partial r^2}+C(r)\frac{\partial h_1}{\partial r}+U(r)h_1=0\ ,
   \label{eq:wave_h1}
\end{equation}
where the coefficients $g(r)$ and $C(r)$ 
depend only on the background solution. 
We study the time evolution of such equation 
adopting a set of lightcone 
coordinates \cite{Gundlach:1993tp,Konoplya:2011qq} 
$u=t-r_*$ and $v=t+r_*$, such that the 
Eq.~\eqref{eq:wave_h1} can be re-written 
as: 
\begin{equation}
\bigg[4\,\frac{\partial^2}{\partial u\, \partial v}+V_a(u,v)\bigg]Z(u,v)=0\ ,
\end{equation}
where $V_a$ reduces to the Regge-Wheeler potential in 
GR in the appropriate limit.
As initial conditions we consider a static Gaussian pulse, \textit{i.e.},
\begin{equation}
    Z(0,v)=A\exp\bigg[-\frac{(v-v_0)^2}{2\sigma^2}\bigg]\, , \; Z(u,0)=0\, .
\end{equation}
We fix the width of the Gaussian pulse to 
$\sigma=5M$ and the constant $v_0=25M$, checking 
that our results are independent of these 
choices.
As the equation is linear, we can freely scale the amplitude to 
$A=1$. The time domain profile of the signal 
is extracted at $r=25M$, setting the integration 
domain to $(u,v)\in [0,400M]$, which is 
large enough to capture the QNM ringing.
Frequencies and damping times of a given 
mode are then computed by numerically 
fitting the signal with damped sinusoids. 
We have carefully chosen the ringdown time 
extraction to avoid any initial data-dependent 
signal or late-time tail behavior, verifying 
that the frequency response and damping time 
is numerically stable by a slight change in 
the ringdown time window.

When solving for the perturbations in the 
frequency domain we first decompose time dependent quantities in Fourier 
modes, such that 
\begin{equation}
h_{1}(t,\omega)=\int_{-\infty}^\infty h_{1}(\omega,r)e^{-i\omega t}d\omega\ .
\end{equation}
The equations for $h_1$ and $h_0$ can then be 
written in a matrix compact form as:
\begin{equation}
    \frac{d}{dr}\boldsymbol{\Psi}_i+C_{ij}\,\boldsymbol{\Psi}_j=0\, ,
\label{eq:wave_Psi}
\end{equation}
where $\boldsymbol{\Psi}= (h_0,h_1)^\top$ and the coefficients $C_{ij}$ depend on the numerical 
background and on $\omega$. 
We solve Eqs.~\eqref{eq:wave_Psi} 
trough the so called matrix method \cite{boyce2004ede}. 
We first impose suitable boundary conditions which 
identify purely ingoing waves at the horizon and 
purely outgoing waves at infinity:
\begin{align}
{\bf \Psi}_i^a(r_\star\rightarrow -\infty)=&\sum_{n_\textnormal{min}}^{n_\textnormal{max}} c_{i}^{(n)}e^{-i\omega r_\star}(r-r_h)^n\label{eq:BChor}\\
{\bf \Psi}_i^a(r_\star\rightarrow+ \infty)=&\sum_{n_\textnormal{min}}^{n_\textnormal{max}} \frac{\bar{c}_{i}^{(n)}}{r^n}e^{i\omega r_\star}\ ,\label{eq:BCinf}
\end{align}
where the coefficients $(c_i^{(n)},\bar{c}_i^{(n)})$ can 
be found with the same iterative procedure adopted 
for Eqs.~\eqref{eq:expA}-\eqref{eq:phi}, and are all 
proportional to the leading term of one of the two metric perturbations.

Let us first examine the behavior of the boundary conditions of the 
perturbations. At the horizon, the leading order 
coefficients in \eqref{eq:BChor} are given by
\begin{align}
    h_0 = & \, c_{1}^{(0)} + \mathcal{O}(r-r_h)\, ,
    \\
    h_1 = & \, \frac{c_{1}^{(0)}}{\sqrt{a^{(1)}b^{(1)}}(r-r_h)}+ \mathcal{O}(1)\, ,
\end{align}
where $c_{1}^{(0)}$ is the leading order amplitude for 
$h_0$, and serves as a free parameter. The expressions 
written above coincide with those obtained in GR.
Deviations enter therefore at the next to leading order 
in the expansion, specifically as:
\begin{equation}
    \frac{h_{0}^{(1)}}{h_{0}^{(0)}}= \frac{N_0}{D_0}\quad , \quad \frac{h_{1}^{(1)}}{h_{0}^{(0)}}= \frac{N_1}{D_1}\, ,
\end{equation}
where
\begin{align}
\begin{split}
    N_{0,1} =\, & f_{\ell,\omega}(a^{(1)},b^{(1)},\phi^{(1)},a^{(2)},b^{(2)},\phi^{(2)},J|_0,\partial_\phi J|_0\\
    & \partial_\phi F|_0,\partial_\phi^2 F|_0)
\end{split}\\
    D_{0,1} =\, & f_{\omega}(a^{(1)},b^{(1)},\phi^{(1)},J|_0,\partial_\phi F|_0)
\end{align}
where in this context $f_{\omega},\,f_{\ell,\omega}$ denote arbitrary functions depending on $\omega$ and $(\ell,\omega)$ respectively.

We fix\footnote{We have 
checked that higher order expansions do not improve our 
numerical calculations.} $n_\text{max}=4$ for the horizon perturbations and $n_\text{max}=7$ at asymptotic infinity. 
Then, we create the $2\times2$ fundamental matrix given 
by two independent solutions of Eq.~\eqref{eq:wave_Psi}:
\begin{equation}
 {\bf X}_a= \big(\boldsymbol{\Psi}^a_{(h)},{\bf \Psi}^a_{(\infty)}\big)
= \begin{pmatrix}
    h_0^{(h)} & h_0^{(\infty)}\\
    h_1^{(h)} & h_1^{(\infty)}\\
 \end{pmatrix}\ .
\end{equation}
The first solution $(h)$ is obtained by integrating 
the differential equations \eqref{eq:wave_Psi} from 
the horizon outward with boundary condition \eqref{eq:BChor}, 
while the second $(\infty)$ is determined by 
integrating from infinity inward using as 
boundary conditions Eqs.~\eqref{eq:BCinf}. 
QNM frequencies correspond to poles of 
the determinant of the fundamental matrix, 
\textit{i.e.} they are given by solving
\begin{equation}
\textnormal{det}\ \textbf{X}(\omega)|_{r_m}=0\, ,
\end{equation}
while we check that our results are independent of the choice for the intermediate matching point radius $r_m$.
%

%
\subsection{The scattering potential}
%
Equation \eqref{eq:wave_h1} allows a straightforward 
check for the stability of the solutions \cite{Kimura:2017uor, Kimura:2018whv, Blazquez-Salcedo:2019nwd, Blazquez-Salcedo:2020rhf}.
To this aim, we first transform $h_1$ in the 
frequency domain, introduce then a generalised tortoise coordinate 
\begin{equation}
    \frac{dr_*}{dr}=g(r)=\sqrt{\frac{1+J-2 (\partial_{\phi }F) B' \phi '-4 B [\phi'(\partial_{\phi }F)]'}{B [A ({J}+1)-2 B (\partial_{\phi }F) A' \phi ']}}\, ,
\end{equation}
and replace $h_1\to f(r)Z$, where $f(r)$ is a 
generic function of the radial coordinate. We 
can then choose $f(r)$ such that Eq.~\eqref{eq:wave_h1} is 
cast into a Regge-Wheeler type form, \textit{i.e.}
\begin{equation}
    \bigg[\frac{d^2}{d r_*^2}+\omega^2\bigg]Z=V_a\, Z\ ,
    \label{eq:h1_pert_ax}
\end{equation}

To assess the stability of the solution we 
then integrate \eqref{eq:h1_pert_ax} as
\begin{equation}
    \int dr_* \omega^2 |Z|^2=\int dr_*\bigg[ \bigg|\frac{dZ}{dr_*}\bigg|^2\hspace{-1mm} +V_a |Z|^2 \bigg]-Z^\star\frac{dZ}{dr_*}\bigg|_{-\infty}^{+\infty},\label{eq:stability1}
\end{equation}
where $\Phi^\star$ is the complex conjugate of 
the perturbation. The last term on the right hand 
side of Eq.~\eqref{eq:stability1} vanishes,
since for unstable modes $Z=0$ both at 
the horizon and at infinity. Then, a sufficient 
condition for the presence of unstable solutions 
is given by
\begin{equation}
    \int_{-\infty}^{+\infty} dr_* V_a < 0\, .
\end{equation}
We can now introduce an arbitrary function $S$, 
that does not diverge at the boundaries, such as to 
re-write Eq.~\eqref{eq:stability1} as 
\begin{equation}
    \int dr_* \omega^2 |Z|^2=\int dr_*\bigg[ \bigg|\frac{dZ}{dr_*}+Z S \bigg|^2 +\hat{V}_a |Z|^2 \bigg]\, ,
\end{equation}
where $\hat{V}_a = V_a + {dS}/{dr_*}-S^2$ is the 
$S$-deformed potential. If the latter is non-negative 
everywhere, then modes are stable. The existence $S$, and hence the stability of the solution, can be 
checked by solving the equation \cite{Kimura:2017uor}:
\begin{equation}
    V_a + \frac{dS}{dr_*}-S^2=0\, .
    \label{eq:S-potential}
\end{equation}

Before closing this subsection let us refer to the 
issue regarding the hyperbolicity of Eq.~\eqref{eq:wave_h1}.
When dealing with the numerical evolution of 
a given set of initial data, we need to be 
particularly cautious about the hyperbolic character 
of the partial differential equations that describe 
the evolution (see for example \cite{Wald:1984rg} for a 
discussion on the problem of well-posedness in GR).
In our case, examining the hyperbolic nature of 
Eq.~\eqref{eq:wave_h1} is fairly straightforward as 
it is determined by the sign of $g(r)$, which 
then depends on the specific model considered. 
Previous calculations showed that, for theories that 
do not feature the scalar-Ricci coupling, 
hyperbolicity may be lost in certain regions of 
the parameter space \cite{Blazquez-Salcedo:2018jnn}.
However, It was later shown in \cite{Antoniou:2022agj} that including the Ricci coupling can render the evolution equations hypebolic at perturbative level. Ref.~\cite{Thaalba:2023fmq} considered spherical collapse and found that this coupling has the same effect at nonlinear level.


%
\section{Numerical results}
\label{sec:results}
%

%
\subsection{A case study}
\label{sec:resultsA}
%

\begin{figure}[!t]
    \centering
    \includegraphics[width=1\linewidth]{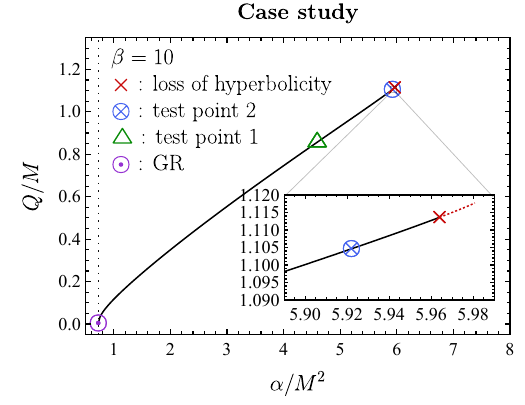}
    \caption{Black hole charge as a function 
    of $\alpha/M^2$ in EsRGB gravity assuming $\beta=10$. Colored markers identify four 
    specific configurations given by the GR limit, 
    a solution close to the parameter 
    space in which the hyperbolicity of Eq.~\eqref{eq:wave_h1} is lost, and two 
    test cases with large values of the 
    charge. Configurations for which hyperbolicity breaks are drawn with 
    a dashed line in the zoomed inset plot. 
    }
    \label{fig:bg}
\end{figure}

\begin{figure*}[!t]
    \centering
    \includegraphics[width=0.49\linewidth]{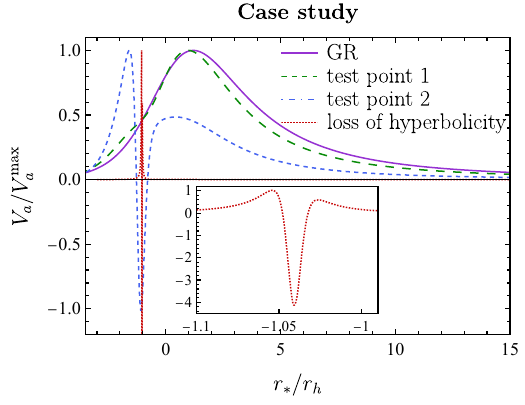}
    \includegraphics[width=0.49\linewidth]{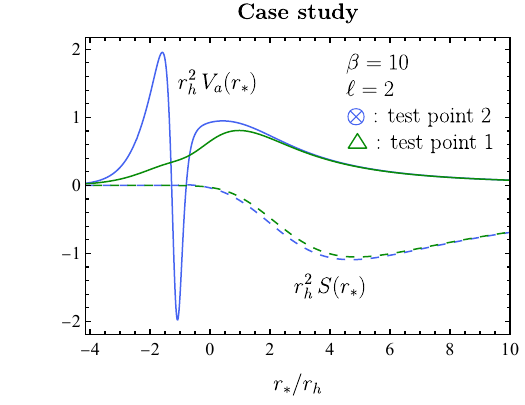}
    \caption{
    \textit{Left:} Axial $\ell=2$ scattering potential 
    for the four BH configurations shown in 
    Fig.~\ref{fig:bg}, normalized by its maximum 
    value. The inset focuses on the critical 
    range where hyperbolicity is lost for the 
    $\times$ model.
    \textit{Right:} Comparison between the $\ell=2$ 
    axial (solid curves) and S-deformed (dashed curves) 
    potentials, for the test points 1 and 2.
    }
    \label{fig:V}
\end{figure*}

\begin{figure}[h]
    \centering
    \includegraphics[width=1\linewidth]{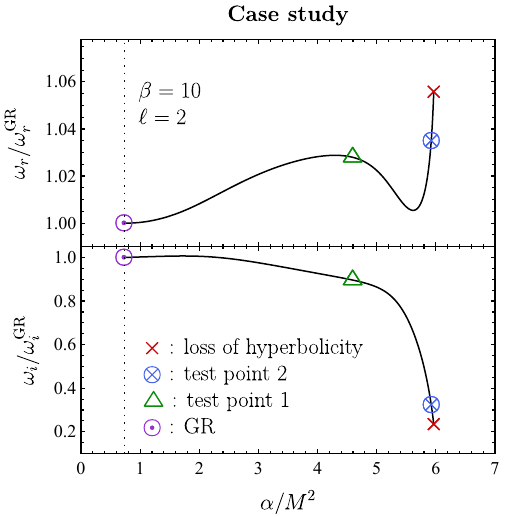}
    \caption{
    Real and imaginary part of the $\ell=2$ 
    mode for BHs in EsRGB gravity with 
    $\beta=10$, as a function of the coupling $\alpha/M^2$. 
    All frequencies are normalised to the 
    Schwarzschild values. On each curve we 
    highlight QNM values for the four 
    specific configurations discussed in 
    Sec.~\ref{sec:resultsA}.
    }
    \label{fig:QNMs}
\end{figure}

To investigate the QNM properties for 
charged BHs we first consider a 
specific case study, focusing on 
EsRGB gravity with $\beta=10$.
Black holes in this theory are radially 
stable, and their charge can be taken to be
large enough to significantly affect the 
background geometry. The charge $Q/M$ of 
such solutions as a function of 
the coupling $\alpha/M^2$ is shown 
in Fig.~\ref{fig:bg}.

We focus on four representative cases 
in the parameter space, with the 
corresponding axial potentials 
drawn in Fig.~\ref{fig:V}.
The first one, just at the onset of the scarization and represented by the label $\odot$, corresponds to a 
Schwarzschild BH in GR and works 
as a comparison model for the 
scalarized BHs. 
The $\triangle$ marker represents a 
configuration with an intermediate 
value of $Q/M$, and features
a scattering potential similar to the 
GR case.
Finally, we consider two highly charged 
solutions, identified 
by $\otimes$ and $\times$, with the latter 
being close to the onset of the hyperbolicity 
loss (see the inset of Fig.~\ref{fig:bg}). 
For the $\otimes$ and $\times$ solutions 
the axial potential becomes negative for 
some values of $r_\star<0$ (see Fig.~\ref{fig:V}), 
affecting the late-time behavior of the perturbations.

Before computing the actual values of the 
QNMs, we analyze the $S$- deformed potential to investigate whether our configurations are stable.
In all cases studied it is possible to find the 
$S$-deformed potential, such that 
Eq.~\eqref{eq:S-potential} is satisfied.
As an example, in the right panel of Fig.~\ref{fig:bg} 
we plot the scattering and the corresponding 
$S$-deformed potential for the $\triangle$ and 
the $\otimes$ test points.

We can now move to study the dependence of the 
QNM frequencies on the BH charge. In Fig.~\ref{fig:QNMs} 
we plot the real and the imaginary part of the 
$\ell=2$ fundamental mode as a function of 
$\alpha/M^2$.
All values are normalised to the corresponding GR 
limit, i.e. $\omega^{\rm GR}_{\ell=2}\approx 0.3737 - i\ 0.08896$. 
The real component of $\omega$ tends to increase 
for larger couplings, varying by less 
than $\sim 8\%$ as $\alpha/M^2$ approaches 
the value at which hyperbolicity is 
broken\footnote{Note that when hyperbolicity 
is lost, modes can still be numerically 
computed}. The imaginary part however 
changes considerably and decreases 
monotonically, varying by up to $80\%$ 
compared to $\omega^{\rm GR}_{i}$. 
This indicates that axial modes  
for scalarized solutions in EsRGB 
gravity can live considerably longer than 
those of Schwarzschild BHs. 

\begin{figure}[!t]
    \centering
    \includegraphics[width=1\linewidth]{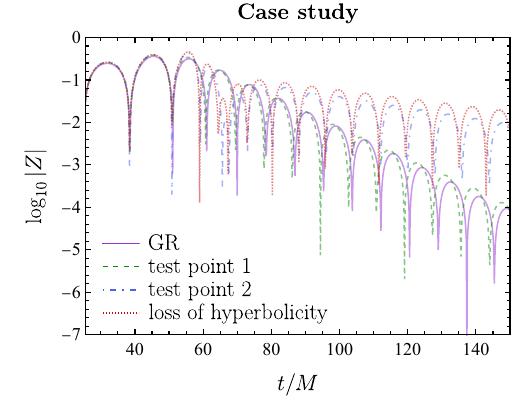}
    \caption{Time evolution of the axial perturbation defined in Eq.~$\eqref{eq:wave_h1}$. 
    Different curves refer to the four 
    BH configurations shown in FIG.~\ref{fig:bg} 
    as colored markers. Perturbations are extracted 
    at $r_\star=25M$.}
    \label{fig:pert}
\end{figure}

Such results are confirmed by a time-evolution study 
of the perturbations, shown in Fig.~\ref{fig:pert} 
for the four BH configurations we focused on, and 
$\ell=2$.
After an initial growth, which depends on the initial 
data, we observe the signal setting into the usual, 
exponentially damped, oscillatory behavior. 
Differences between the four models are 
mainly controlled by changes in the scattering 
potential. As expected, perturbations for the 
$\otimes$ and $\times$ models show a slower decay. 
We have numerically extracted the values of the 
frequency and of the damping time for each perturbation, 
finding good agreement with the results obtained with 
the frequency domain analysis.

%
\subsection{Potential and quasinormal modes}
%

\begin{figure}[!t]
    \centering
    \includegraphics[width=1\linewidth]{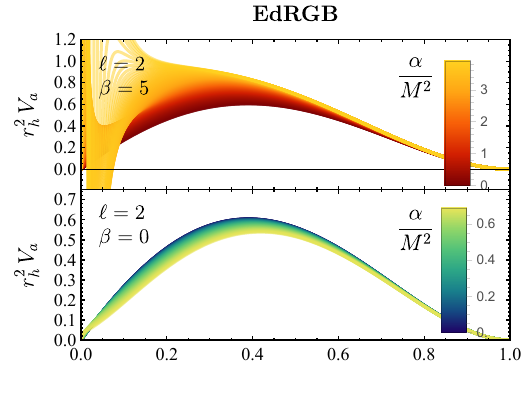}\\[-5mm]
    \includegraphics[width=1\linewidth]{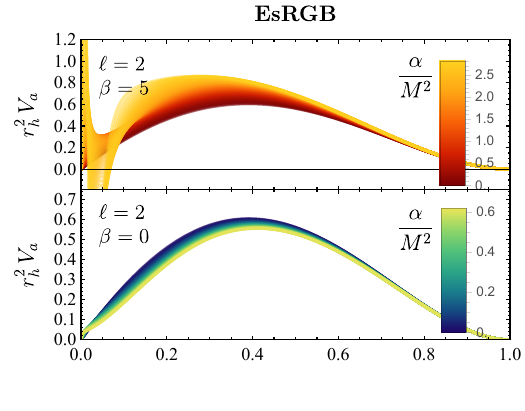}\\[-5mm]
    \includegraphics[width=1\linewidth]{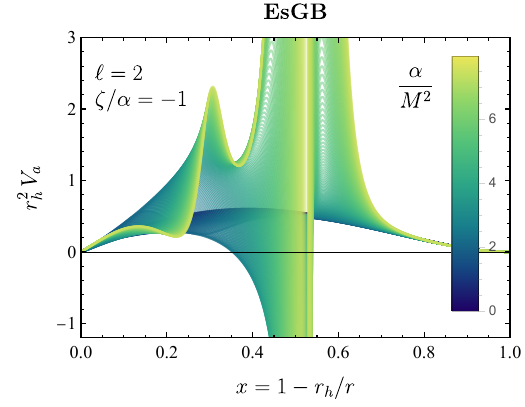}
    \caption{
    \textit{Top:}
    Axial potential in the EdRGB model for $\ell=2$ and $\beta=0,\,2$. Each color in the bar legend corresponds to a different potential curve created for a different value of the dimensionless coupling $\alpha/M^2$, spanning all the allowed values.
    \textit{Middle:} Same but for the EsGB model. Here the nonzero value for the Ricci-scalar coupling is chosen to be $\beta=5$.
    \textit{Bottom:} Similar plot but for the EsGB model when $\zeta/\alpha=-1,\,-3/2$.
    }
    \label{fig:potential}
\end{figure}

Having used the special case $\beta=10$ in the EsRGB scenario as a demonstration of how our analysis proceeds, we now broaden the discussion to the axial potential and QNMs concerning the EdRGB, EsRGB, and EsGB theories as those were introduced in Sec.~\ref{sec:theory}.

In the three panels included in Fig.~\ref{fig:potential} we plot the axial potential for angular number $\ell=2$ in the three different theories.
As has already been explained, after fixing the parameters $\beta,\, \zeta$, there is a range of values for the GB coupling $\alpha$ that allows for asymptotically flat black holes with a vanishing scalar field, which explains the different values of $\alpha$ explored in each theory subset in Fig.~\ref{fig:potential}.
In principle, the larger $\alpha/M^2$ is, the more the black hole deviates from GR.
For EdRGB and EsRGB, we plot the potential for $\beta=0$ and $\beta=5$. For EsGB we only plot the potential for $\zeta/\alpha=-1$, since in the limit $\zeta/\alpha\rightarrow 0$ the theory coincides with EsRGB for $\beta\rightarrow 0$.

We see that when $\beta=0$ in EsRGB and EsGB, or $\zeta/\alpha=0$ in EsGB, the potential remains positive and relatively unchanged along the allowed range for $\alpha/M^2$.
Introducing the additional couplings ($\beta,\, \zeta$) significantly affects the form of the potential.
In particular, in EdRGB and EsRGB the potential develops an increasingly negative region when one moves along the $\alpha/M^2$ ``existence lines''.
In the EsGB case for a negative ratio $\zeta/\alpha$, and as one moves along the existence line towards larger values for $\alpha/M^2$, the potential develops local minima in addition to the one corresponding to the negative region.

\begin{figure}[!t]
    \centering
    \includegraphics[width=1\linewidth]{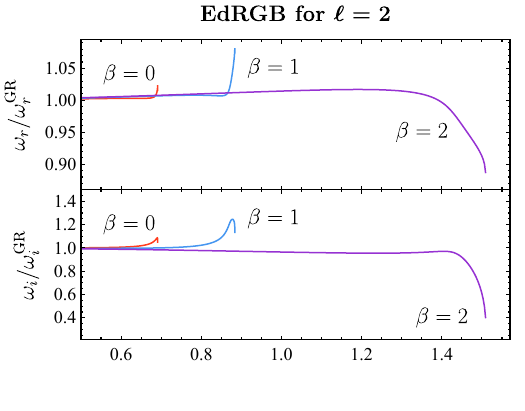}\\[-5mm]
    \includegraphics[width=1\linewidth]{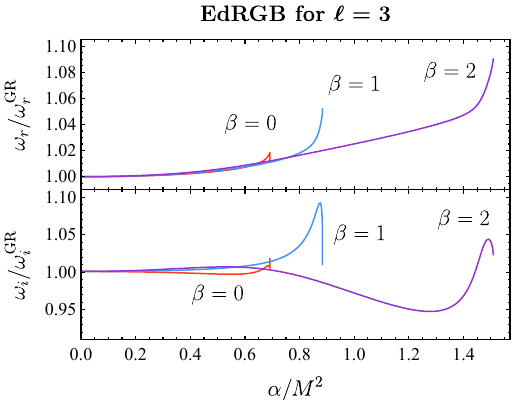}
    \caption{
    \textit{Top:} Relative ratio of the hairy black hole mode with respect to the GR one, in the EdRGB model for $\ell=2$ and $\beta=0,\,1,\, 2$.
    \textit{Bottom:} Same as top panel but for $\ell=3$.
    }
    \label{fig:EdRGB_axial_modes}
\end{figure}
\begin{figure}[!t]
    \centering
    \includegraphics[width=1\linewidth]{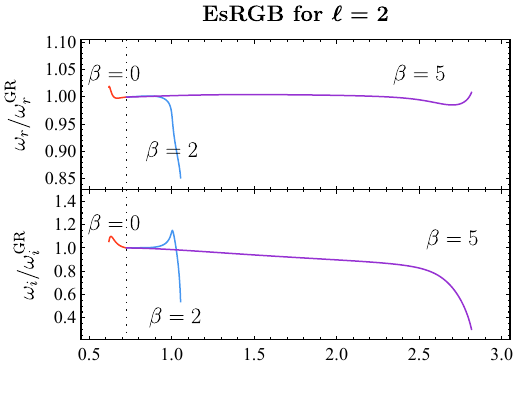}\\[-5mm]
    \includegraphics[width=1\linewidth]{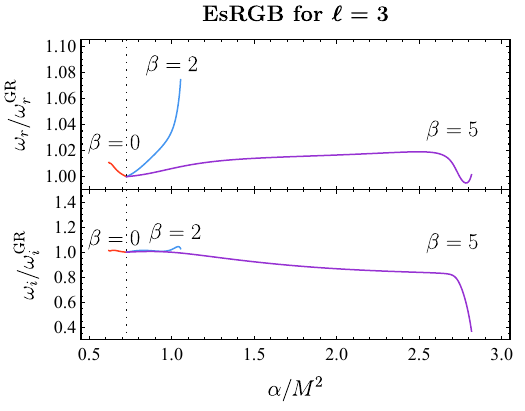}
    \caption{
    \textit{Top:} Relative ratio of the hairy black hole mode with respect to the GR one, in the EsRGB model for $\ell=2$ and $\beta=0,\,2,\, 5$.
    \textit{Bottom:} Same as top panel but for $\ell=3$.
    }
    \label{fig:EsRGB_axial_modes}
\end{figure}
\begin{figure}[!t]
    \centering
    \includegraphics[width=1\linewidth]{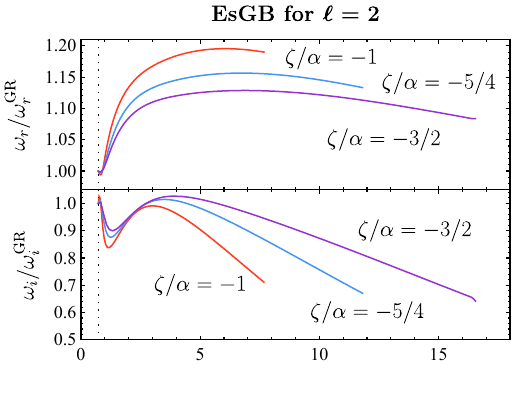}\\[-5mm]
    \includegraphics[width=1\linewidth]{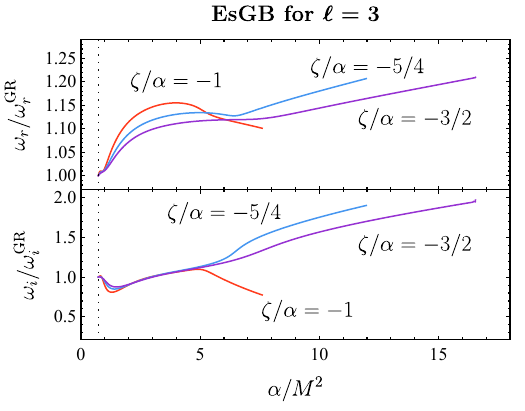}
    \caption{
    \textit{Top:
    } Relative ratio of the hairy black hole mode with respect to the GR one, in the EsGB model for $\ell=2$ and $\zeta/\alpha=-1,\,-5/4,\,-3/2$.
    \textit{Bottom:} Same as top panel but for $\ell=3$.
    }
    \label{fig:EsGB_axial_modes}
\end{figure}

We then move on to calculating the QNM frequencies according to the integration method we described in Sec.~\ref{sec:bg_and_pert}.
In Fig.~\ref{fig:EdRGB_axial_modes}, we present the results in the EdRGB scenario.
For $\beta=0$ and $\ell=2$, the results are in agreement with those recovered in \cite{Blazquez-Salcedo:2016enn}.
Increasing the values of $\beta$, results in increasingly larger deviations from GR particularly towards the end of the existence line, where the ratio $\alpha/M^2$ is maximized.
In accordance to the form of the effective potential for large $\alpha/M^2$, which showed the appearance of a negative region, the ratio of the imaginary part of the QNM frequencies tends to smaller positive values, but never reaches zero.
It is worth pointing out that according to the S-deformed potential method, no unstable modes are found, consistently with the aforementioned results.
It is also worth-noting that for $\ell=2$, while both the real and imaginary parts increase towards the end of the existence line for smaller values of $\beta$, for larger $\beta$ the opposite behaviour is observed.
Especially the imaginary component for $\beta=2$ drops by as much as $\sim 60\%$ near the edge.
The real part, even though deviates significantly more than what it does in the $\beta=0$ scenario, it remains within around $\sim 10\%$ of its GR counterpart in the $\beta=2$ case.
For $\ell=3$, on the other hand, the behaviour of the QNM curves is different, \textit{e.g.} the imaginary part for $\beta=1,\,2$ is significantly non-monotonic along the existence lines.

Similar-scale results are depicted in Fig.~\ref{fig:EsRGB_axial_modes} concerning the EsRGB model, where the vertical dashed line corresponds to the $n=0$ scalarization threshold.
Of particular interest in this model is the $\beta=2$ case for which significant deviations from GR are noticed relatively close to the scalarization threshold, as is evident especially for the real part of the $\ell=3$ mode.
Once again, in agreement with the potential presented in the middle panel of Fig.~\ref{fig:potential}, the imaginary part of the QNM frequency for $\ell=2$, becomes all the less negative, as one moves along the existence lines, but never reaches the horizontal axis.

For the EsGB case, the results are shown in Fig.~\ref{fig:EsGB_axial_modes}, and between the three models, they are the ones that deviate from GR the most.
It is worth emphasizing that in this model significant deviations from GR appear fairly close to the scalarization threshold regardless of the value of the ratio $\zeta/\alpha$.

%
\section{Conclusions}
\label{sec:conclusions}
%

We have studied axial quasinormal modes of black holes with scalar hair in several models that evade no-hair theorems and, in some cases, exhibit spontaneous scalarization. We analysed perturbations in both the time and the frequency domain, extending previous studies in the literature. 

In particular, we have shown that in the minimal case where the scalar couples only to the Gauss-Bonnet invariant, the scalar charge has a small impact on the axial perturbations, deviating only slightly from the Schwarzschild BH in GR. However, when a coupling to the Ricci scalar is added some configurations can vibrate for a significantly longer time, even though the oscillation frequency is just a few percent off from its GR counterpart.

It is worth pointing out the following. First, for the EdRGB and EsRGB theories the maximum deviation of the modes from the GR case is for smaller masses and close to the threshold mass beyond which black holes cease to exist for a given value of the coupling. This is not necessarily the case for the quadratic EsGB model.
Second, the maximum deviations from GR appear in the parameter space region where the background solutions' scalar charge is also maximum. This is expected, as the charge is usually related to the intensity of the additional terms appearing in the perturbed equations.

Our analysis showed that the effective potential of axial perturbations can become negative close to the BH as we increase the coupling parameter of the additional interactions and also develop a well and an additional minimum. The potential well could indicate the existence of trapped unstable modes. However, studying the deformed potential,  performing a direct integration mode computation, and performing a time-domain evolution of initial signals, we found no evidence of a mode instability. Moreover, we have also not found any discontinuous behavior of the fundamental mode and only moderate changes in the QNM frequencies. This suggests that the axial spectra of the BHs studied here are stable. This is to be contrasted with  scenarios where small (ad hoc) deviations in the potential might lead to a spectral instability of the modes~\cite{Cheung:2021bol}.

Our analysis is restricted to the axial sector. The polar sector is more evolved as gravitational and scalar perturbations are coupled to each other, with the scalar sector leaving a possible imprint in the ringdown. Usually, the polar sector dominates over the axial one, but in the possibility of the imaginary part of the axial modes being smaller than the polar in generalized scalar-tensor theories, this hierarchy could be broken. The axial modes could live longer than the polar ones. However, only a full computation of the spectrum would reveal the hierarchy, which is a natural extension of this work. 

The lifetimes of axial and polar modes 
play a crucial role on the detectability 
of GR deviations which, for theories of gravity with dimensionful couplings, represent a difficult challenge. 
A recent analysis for EdGB dilatonic BHs suggested that, even with next 
generation detectors, ringdown observations alone may not be 
sufficient to measure the coupling of the theory 
\cite{Maselli:2023khq}, since constrains are dominated by 
light objects with low signal to noise ratios. 
Our results, however, show that deviations in the QNM 
spectrum increase for scalarized BHs, leading in particular 
to modes with significant longer damping times.  
We plan to assess the ability of 3g  interferometers to 
detect such changes having computed the full 
axial and polar sector.

\begin{acknowledgments}
C.F.B.M. would like
to thank Fundação Amazônia de Amparo a Estudos
e Pesquisas (FAPESPA), Conselho Nacional de Desenvolvimento Científico e Tecnológico (CNPq) and Coordenação de Aperfeiçoamento de Pessoal de Nível Superior (CAPES) – Finance Code 001, from Brazil, for partial financial support. 
A. M. acknowledges partial support from 
MUR PRIN Grant No. 2022-Z9X4XS, funded by the European 
Union - Next Generation EU.
T.P.S. acknowledges partial support from
the STFC Consolidated Grants no.~ST/X000672/1 and
no.~ST/V005596/1. 

\end{acknowledgments}

\newpage

\appendix

%
\section{Background equations}
\label{ap:background}
%

Here we provide the explicit form of the background equations \eqref{eq:bg_grav} and \eqref{eq:bg_scalar} in the metric ansantz \eqref{eq:metric}.
From the nontrivial components of the Einstein field equations at zeroth order and the background equation of motion for the scalar field we find
\begin{align}
        (tt):\quad
        &
        4+2 r^2 V-8 \partial _{\phi }F B' \phi '+16 B^2 (\partial _{\phi }F \phi ''\nonumber\\
        &
        +\partial _{\phi }{}^2F (\phi ')^2)-B (-24 \partial _{\phi }F B' \phi '+4 J\nonumber\\
        &
        +4 r^2 \partial _{\phi }J \phi ''+4 r^2 \partial_{\phi }^2J (\phi ')^2+r^2 (\phi')^2\\
        &
        +8 r \partial _{\phi }J \phi '+16 \partial_{\phi }F \phi ''+16 \partial _{\phi }{}^2F (\phi ')^2
        \nonumber\\
        &
        +4)-4 J \left(r B'-1\right)-2 r^2 \partial _{\phi }J B' \phi'\nonumber\\
        &
        -4 r B'=0\, ,\nonumber\\[2mm]
        (rr):\quad
        &
        4 {J}+-2 B A' (4 (3 B-1) \partial _{\phi }F \phi '-2 {J} r
        \nonumber\\
        &
        -r (r \partial _{\phi }J \phi '+2))-A (B (-4 {J}+r^2 (\phi')^2\\
        &
        -8 r \partial _{\phi }J \phi '-4)+2 r^2 V+4)=0\, ,\nonumber\\[2mm]
        (\phi):\quad
        &
        A (A (r^2 (B' \phi '+2 V'(\phi ))+\partial _{\phi }J (4-4 r B'))\nonumber\\
        &
        -A' B' (r^2 \partial _{\phi }J+4 \partial _{\phi }F))-4 B^2 \partial _{\phi }F ((A')^2
        \nonumber\\
        &
        -2 A A'')+B (-2 A
        (A'' (r^2 \partial _{\phi }J+4 \partial _{\phi }F)\\
        &
        +A (2 \partial _{\phi }J-r (r \phi ''+2 \phi ')))\nonumber\\
        &
        +A A' (12 \partial _{\phi }F B'+r (r \phi '-4 \partial _{\phi }J))\nonumber\\
        &
        +(A')^2 (r^2 \partial _{\phi }J+4 \partial _{\phi }F))=0\, .\nonumber
\end{align}
%

%
\section{Perturbation equations}
\label{ap:perturbation_equations}
%
In this appendix we provide the explicit form of 
the ODEs for the axial perturbations $(h_0(r),h_1(r))$ 
in Eq.~\eqref{eq:axial_RW_gauge}:
\begin{align}
           (r\chi):\quad
           &
           h_0'-\frac{2 h_0}{r}+i h_1 \{A^2 [2 J (r B'+\ell ^2+\ell -2)\nonumber\\
           &
           +2 r^2 (\partial_{\phi }J) B' \phi '+2 r B'+4 B r^2 (\partial_{\phi}J) \phi''\nonumber\\
           &
           +4 B r^2 (\partial_{\phi}^2J) \phi'^2
           +B r^2 \phi'^2+4 B r (\partial _{\phi }J) \phi'
           \nonumber\\
           &
           -2 r^2 V(\phi )+2 \ell ^2+2 \ell -4]\nonumber\\
           &
           +B r A'^2 (4 B (\partial _{\phi }F) \phi '-J r-r)\nonumber\\
           &
           -8 B^2 r ((\partial _{\phi }^2 F) A' \phi'^2
           +A [r^2 (A' B'-2 \omega ^2)\label{eq:r-chi}\\
           &
           +(\partial _{\phi }F) (A'' \phi '+A' \phi ''))+J r (r A' B'
           \nonumber\\
           &
           +2 B (r A''+A')-2 r \omega ^2)+2 B (r (r A''
           \nonumber\\
           &
           +4 (\partial_{\phi}F) \omega ^2 \phi ')+A' (r (r (\partial _{\phi }J) \phi '+1)
           \nonumber\\
           &
           -2 (\partial _{\phi }F) \phi '
           (3 r B'+\ell ^2+\ell -2)))]\}\nonumber\\
           &
           \big/
           2 A r \omega  (-J r+4 B (\partial_{\phi }F) \phi '-r)
           =0\, ,\nonumber\\[2mm]
           (\theta\chi):\quad
           &
           h_1'-h_0,i \omega  [2 (\partial_{\phi}F) B' \phi '+4 B (\partial_{\phi}F) \phi ''\nonumber\\
           &
           +4 B (\partial_{\phi}^2F) \phi'^2-J-1]/\{B [A (J+1)\nonumber\\
           &
           -2 B (\partial_{\phi}F) A' \phi']\}
           +h_1\{[A^2 ((J+1) B'\label{eq:theta-chi}\\
           &
           +2 B (\partial_{\phi}J) \phi')]
           +2 B^2 (\partial_{\phi }F) A'^2 \phi '\nonumber\\
           &
           -A B [4 B (\partial_{\phi}F) A'' \phi '+A' (6 (\partial_{\phi }F) B' \phi'\nonumber\\
           &
           +4 B (\partial_{\phi }F) \phi ''+4 B (\partial_{\phi}^2F) \phi'^2-J-1)]\}\nonumber\\
           &
           /[2 A B A (J+1)-2 B (\partial_{\phi}F) A' \phi')]
           =0 \, .\nonumber
\end{align}

\bibliography{bibnote}

\begin{thebibliography}{65}%
\makeatletter
\providecommand \@ifxundefined [1]{%
 \@ifx{#1\undefined}
}%
\providecommand \@ifnum [1]{%
 \ifnum #1\expandafter \@firstoftwo
 \else \expandafter \@secondoftwo
 \fi
}%
\providecommand \@ifx [1]{%
 \ifx #1\expandafter \@firstoftwo
 \else \expandafter \@secondoftwo
 \fi
}%
\providecommand \natexlab [1]{#1}%
\providecommand \enquote  [1]{``#1''}%
\providecommand \bibnamefont  [1]{#1}%
\providecommand \bibfnamefont [1]{#1}%
\providecommand \citenamefont [1]{#1}%
\providecommand \href@noop [0]{\@secondoftwo}%
\providecommand \href [0]{\begingroup \@sanitize@url \@href}%
\providecommand \@href[1]{\@@startlink{#1}\@@href}%
\providecommand \@@href[1]{\endgroup#1\@@endlink}%
\providecommand \@sanitize@url [0]{\catcode `\\12\catcode `\$12\catcode
  `\&12\catcode `\#12\catcode `\^12\catcode `\_12\catcode `\%12\relax}%
\providecommand \@@startlink[1]{}%
\providecommand \@@endlink[0]{}%
\providecommand \url  [0]{\begingroup\@sanitize@url \@url }%
\providecommand \@url [1]{\endgroup\@href {#1}{\urlprefix }}%
\providecommand \urlprefix  [0]{URL }%
\providecommand \Eprint [0]{\href }%
\providecommand \doibase [0]{http://dx.doi.org/}%
\providecommand \selectlanguage [0]{\@gobble}%
\providecommand \bibinfo  [0]{\@secondoftwo}%
\providecommand \bibfield  [0]{\@secondoftwo}%
\providecommand \translation [1]{[#1]}%
\providecommand \BibitemOpen [0]{}%
\providecommand \bibitemStop [0]{}%
\providecommand \bibitemNoStop [0]{.\EOS\space}%
\providecommand \EOS [0]{\spacefactor3000\relax}%
\providecommand \BibitemShut  [1]{\csname bibitem#1\endcsname}%
\let\auto@bib@innerbib\@empty
\bibitem [{\citenamefont {Abbott}\ \emph {et~al.}(2016)\citenamefont {Abbott}
  \emph {et~al.}}]{LIGOScientific:2016aoc}%
  \BibitemOpen
  \bibfield  {author} {\bibinfo {author} {\bibfnamefont {B.~P.}\ \bibnamefont
  {Abbott}} \emph {et~al.} (\bibinfo {collaboration} {LIGO Scientific,
  Virgo}),\ }\href {\doibase 10.1103/PhysRevLett.116.061102} {\bibfield
  {journal} {\bibinfo  {journal} {Phys. Rev. Lett.}\ }\textbf {\bibinfo
  {volume} {116}},\ \bibinfo {pages} {061102} (\bibinfo {year} {2016})},\
  \Eprint {http://arxiv.org/abs/1602.03837} {arXiv:1602.03837 [gr-qc]}
  \BibitemShut {NoStop}%
\bibitem [{\citenamefont {Abbott}\ \emph {et~al.}(2019)\citenamefont {Abbott}
  \emph {et~al.}}]{LIGOScientific:2018mvr}%
  \BibitemOpen
  \bibfield  {author} {\bibinfo {author} {\bibfnamefont {B.~P.}\ \bibnamefont
  {Abbott}} \emph {et~al.} (\bibinfo {collaboration} {LIGO Scientific,
  Virgo}),\ }\href {\doibase 10.1103/PhysRevX.9.031040} {\bibfield  {journal}
  {\bibinfo  {journal} {Phys. Rev. X}\ }\textbf {\bibinfo {volume} {9}},\
  \bibinfo {pages} {031040} (\bibinfo {year} {2019})},\ \Eprint
  {http://arxiv.org/abs/1811.12907} {arXiv:1811.12907 [astro-ph.HE]}
  \BibitemShut {NoStop}%
\bibitem [{\citenamefont {Abbott}\ \emph
  {et~al.}(2021{\natexlab{a}})\citenamefont {Abbott} \emph
  {et~al.}}]{LIGOScientific:2020ibl}%
  \BibitemOpen
  \bibfield  {author} {\bibinfo {author} {\bibfnamefont {R.}~\bibnamefont
  {Abbott}} \emph {et~al.} (\bibinfo {collaboration} {LIGO Scientific,
  Virgo}),\ }\href {\doibase 10.1103/PhysRevX.11.021053} {\bibfield  {journal}
  {\bibinfo  {journal} {Phys. Rev. X}\ }\textbf {\bibinfo {volume} {11}},\
  \bibinfo {pages} {021053} (\bibinfo {year} {2021}{\natexlab{a}})},\ \Eprint
  {http://arxiv.org/abs/2010.14527} {arXiv:2010.14527 [gr-qc]} \BibitemShut
  {NoStop}%
\bibitem [{\citenamefont {Abbott}\ \emph
  {et~al.}(2021{\natexlab{b}})\citenamefont {Abbott} \emph
  {et~al.}}]{LIGOScientific:2021usb}%
  \BibitemOpen
  \bibfield  {author} {\bibinfo {author} {\bibfnamefont {R.}~\bibnamefont
  {Abbott}} \emph {et~al.} (\bibinfo {collaboration} {LIGO Scientific,
  VIRGO}),\ }\href@noop {} {\  (\bibinfo {year} {2021}{\natexlab{b}})},\
  \Eprint {http://arxiv.org/abs/2108.01045} {arXiv:2108.01045 [gr-qc]}
  \BibitemShut {NoStop}%
\bibitem [{\citenamefont {Abbott}\ \emph
  {et~al.}(2021{\natexlab{c}})\citenamefont {Abbott} \emph
  {et~al.}}]{LIGOScientific:2021djp}%
  \BibitemOpen
  \bibfield  {author} {\bibinfo {author} {\bibfnamefont {R.}~\bibnamefont
  {Abbott}} \emph {et~al.} (\bibinfo {collaboration} {LIGO Scientific, VIRGO,
  KAGRA}),\ }\href@noop {} {\  (\bibinfo {year} {2021}{\natexlab{c}})},\
  \Eprint {http://arxiv.org/abs/2111.03606} {arXiv:2111.03606 [gr-qc]}
  \BibitemShut {NoStop}%
\bibitem [{\citenamefont {Horndeski}(1974)}]{Horndeski:1974wa}%
  \BibitemOpen
  \bibfield  {author} {\bibinfo {author} {\bibfnamefont {G.~W.}\ \bibnamefont
  {Horndeski}},\ }\href {\doibase 10.1007/BF01807638} {\bibfield  {journal}
  {\bibinfo  {journal} {Int. J. Theor. Phys.}\ }\textbf {\bibinfo {volume}
  {10}},\ \bibinfo {pages} {363} (\bibinfo {year} {1974})}\BibitemShut
  {NoStop}%
\bibitem [{\citenamefont {Deffayet}\ \emph {et~al.}(2009)\citenamefont
  {Deffayet}, \citenamefont {Deser},\ and\ \citenamefont
  {Esposito-Farese}}]{Deffayet:2009mn}%
  \BibitemOpen
  \bibfield  {author} {\bibinfo {author} {\bibfnamefont {C.}~\bibnamefont
  {Deffayet}}, \bibinfo {author} {\bibfnamefont {S.}~\bibnamefont {Deser}}, \
  and\ \bibinfo {author} {\bibfnamefont {G.}~\bibnamefont {Esposito-Farese}},\
  }\href {\doibase 10.1103/PhysRevD.80.064015} {\bibfield  {journal} {\bibinfo
  {journal} {Phys. Rev. D}\ }\textbf {\bibinfo {volume} {80}},\ \bibinfo
  {pages} {064015} (\bibinfo {year} {2009})},\ \Eprint
  {http://arxiv.org/abs/0906.1967} {arXiv:0906.1967 [gr-qc]} \BibitemShut
  {NoStop}%
\bibitem [{\citenamefont {Kobayashi}\ \emph {et~al.}(2011)\citenamefont
  {Kobayashi}, \citenamefont {Yamaguchi},\ and\ \citenamefont
  {Yokoyama}}]{Kobayashi:2011nu}%
  \BibitemOpen
  \bibfield  {author} {\bibinfo {author} {\bibfnamefont {T.}~\bibnamefont
  {Kobayashi}}, \bibinfo {author} {\bibfnamefont {M.}~\bibnamefont
  {Yamaguchi}}, \ and\ \bibinfo {author} {\bibfnamefont {J.}~\bibnamefont
  {Yokoyama}},\ }\href {\doibase 10.1143/PTP.126.511} {\bibfield  {journal}
  {\bibinfo  {journal} {Prog. Theor. Phys.}\ }\textbf {\bibinfo {volume}
  {126}},\ \bibinfo {pages} {511} (\bibinfo {year} {2011})},\ \Eprint
  {http://arxiv.org/abs/1105.5723} {arXiv:1105.5723 [hep-th]} \BibitemShut
  {NoStop}%
\bibitem [{\citenamefont {Elder}\ and\ \citenamefont
  {Sakstein}(2023)}]{Elder:2022rak}%
  \BibitemOpen
  \bibfield  {author} {\bibinfo {author} {\bibfnamefont {B.}~\bibnamefont
  {Elder}}\ and\ \bibinfo {author} {\bibfnamefont {J.}~\bibnamefont
  {Sakstein}},\ }\href {\doibase 10.1103/PhysRevD.107.044006} {\bibfield
  {journal} {\bibinfo  {journal} {Phys. Rev. D}\ }\textbf {\bibinfo {volume}
  {107}},\ \bibinfo {pages} {044006} (\bibinfo {year} {2023})},\ \Eprint
  {http://arxiv.org/abs/2210.10955} {arXiv:2210.10955 [gr-qc]} \BibitemShut
  {NoStop}%
\bibitem [{\citenamefont {Bekenstein}(1995)}]{Bekenstein:1995un}%
  \BibitemOpen
  \bibfield  {author} {\bibinfo {author} {\bibfnamefont {J.~D.}\ \bibnamefont
  {Bekenstein}},\ }\href {\doibase 10.1103/PhysRevD.51.R6608} {\bibfield
  {journal} {\bibinfo  {journal} {Phys. Rev. D}\ }\textbf {\bibinfo {volume}
  {51}},\ \bibinfo {pages} {R6608} (\bibinfo {year} {1995})}\BibitemShut
  {NoStop}%
\bibitem [{\citenamefont {Hawking}(1972)}]{Hawking:1972qk}%
  \BibitemOpen
  \bibfield  {author} {\bibinfo {author} {\bibfnamefont {S.~W.}\ \bibnamefont
  {Hawking}},\ }\href {\doibase 10.1007/BF01877518} {\bibfield  {journal}
  {\bibinfo  {journal} {Commun. Math. Phys.}\ }\textbf {\bibinfo {volume}
  {25}},\ \bibinfo {pages} {167} (\bibinfo {year} {1972})}\BibitemShut
  {NoStop}%
\bibitem [{\citenamefont {Sotiriou}\ and\ \citenamefont
  {Faraoni}(2012)}]{Sotiriou:2011dz}%
  \BibitemOpen
  \bibfield  {author} {\bibinfo {author} {\bibfnamefont {T.~P.}\ \bibnamefont
  {Sotiriou}}\ and\ \bibinfo {author} {\bibfnamefont {V.}~\bibnamefont
  {Faraoni}},\ }\href {\doibase 10.1103/PhysRevLett.108.081103} {\bibfield
  {journal} {\bibinfo  {journal} {Phys. Rev. Lett.}\ }\textbf {\bibinfo
  {volume} {108}},\ \bibinfo {pages} {081103} (\bibinfo {year} {2012})},\
  \Eprint {http://arxiv.org/abs/1109.6324} {arXiv:1109.6324 [gr-qc]}
  \BibitemShut {NoStop}%
\bibitem [{\citenamefont {Silva}\ \emph {et~al.}(2018)\citenamefont {Silva},
  \citenamefont {Sakstein}, \citenamefont {Gualtieri}, \citenamefont
  {Sotiriou},\ and\ \citenamefont {Berti}}]{Silva:2017uqg}%
  \BibitemOpen
  \bibfield  {author} {\bibinfo {author} {\bibfnamefont {H.~O.}\ \bibnamefont
  {Silva}}, \bibinfo {author} {\bibfnamefont {J.}~\bibnamefont {Sakstein}},
  \bibinfo {author} {\bibfnamefont {L.}~\bibnamefont {Gualtieri}}, \bibinfo
  {author} {\bibfnamefont {T.~P.}\ \bibnamefont {Sotiriou}}, \ and\ \bibinfo
  {author} {\bibfnamefont {E.}~\bibnamefont {Berti}},\ }\href {\doibase
  10.1103/PhysRevLett.120.131104} {\bibfield  {journal} {\bibinfo  {journal}
  {Phys. Rev. Lett.}\ }\textbf {\bibinfo {volume} {120}},\ \bibinfo {pages}
  {131104} (\bibinfo {year} {2018})},\ \Eprint
  {http://arxiv.org/abs/1711.02080} {arXiv:1711.02080 [gr-qc]} \BibitemShut
  {NoStop}%
\bibitem [{\citenamefont {Mignemi}\ and\ \citenamefont
  {Stewart}(1993)}]{Mignemi:1992pm}%
  \BibitemOpen
  \bibfield  {author} {\bibinfo {author} {\bibfnamefont {S.}~\bibnamefont
  {Mignemi}}\ and\ \bibinfo {author} {\bibfnamefont {N.~R.}\ \bibnamefont
  {Stewart}},\ }\href {\doibase 10.1016/0370-2693(93)91824-7} {\bibfield
  {journal} {\bibinfo  {journal} {Phys. Lett. B}\ }\textbf {\bibinfo {volume}
  {298}},\ \bibinfo {pages} {299} (\bibinfo {year} {1993})},\ \Eprint
  {http://arxiv.org/abs/hep-th/9206018} {arXiv:hep-th/9206018} \BibitemShut
  {NoStop}%
\bibitem [{\citenamefont {Kanti}\ \emph {et~al.}(1996)\citenamefont {Kanti},
  \citenamefont {Mavromatos}, \citenamefont {Rizos}, \citenamefont {Tamvakis},\
  and\ \citenamefont {Winstanley}}]{Kanti:1995vq}%
  \BibitemOpen
  \bibfield  {author} {\bibinfo {author} {\bibfnamefont {P.}~\bibnamefont
  {Kanti}}, \bibinfo {author} {\bibfnamefont {N.~E.}\ \bibnamefont
  {Mavromatos}}, \bibinfo {author} {\bibfnamefont {J.}~\bibnamefont {Rizos}},
  \bibinfo {author} {\bibfnamefont {K.}~\bibnamefont {Tamvakis}}, \ and\
  \bibinfo {author} {\bibfnamefont {E.}~\bibnamefont {Winstanley}},\ }\href
  {\doibase 10.1103/PhysRevD.54.5049} {\bibfield  {journal} {\bibinfo
  {journal} {Phys. Rev. D}\ }\textbf {\bibinfo {volume} {54}},\ \bibinfo
  {pages} {5049} (\bibinfo {year} {1996})},\ \Eprint
  {http://arxiv.org/abs/hep-th/9511071} {arXiv:hep-th/9511071} \BibitemShut
  {NoStop}%
\bibitem [{\citenamefont {Sotiriou}\ and\ \citenamefont
  {Zhou}(2014{\natexlab{a}})}]{Sotiriou:2013qea}%
  \BibitemOpen
  \bibfield  {author} {\bibinfo {author} {\bibfnamefont {T.~P.}\ \bibnamefont
  {Sotiriou}}\ and\ \bibinfo {author} {\bibfnamefont {S.-Y.}\ \bibnamefont
  {Zhou}},\ }\href {\doibase 10.1103/PhysRevLett.112.251102} {\bibfield
  {journal} {\bibinfo  {journal} {Phys. Rev. Lett.}\ }\textbf {\bibinfo
  {volume} {112}},\ \bibinfo {pages} {251102} (\bibinfo {year}
  {2014}{\natexlab{a}})},\ \Eprint {http://arxiv.org/abs/1312.3622}
  {arXiv:1312.3622 [gr-qc]} \BibitemShut {NoStop}%
\bibitem [{\citenamefont {Antoniou}\ \emph
  {et~al.}(2018{\natexlab{a}})\citenamefont {Antoniou}, \citenamefont
  {Bakopoulos},\ and\ \citenamefont {Kanti}}]{Antoniou:2017acq}%
  \BibitemOpen
  \bibfield  {author} {\bibinfo {author} {\bibfnamefont {G.}~\bibnamefont
  {Antoniou}}, \bibinfo {author} {\bibfnamefont {A.}~\bibnamefont
  {Bakopoulos}}, \ and\ \bibinfo {author} {\bibfnamefont {P.}~\bibnamefont
  {Kanti}},\ }\href {\doibase 10.1103/PhysRevLett.120.131102} {\bibfield
  {journal} {\bibinfo  {journal} {Phys. Rev. Lett.}\ }\textbf {\bibinfo
  {volume} {120}},\ \bibinfo {pages} {131102} (\bibinfo {year}
  {2018}{\natexlab{a}})},\ \Eprint {http://arxiv.org/abs/1711.03390}
  {arXiv:1711.03390 [hep-th]} \BibitemShut {NoStop}%
\bibitem [{\citenamefont {Doneva}\ and\ \citenamefont
  {Yazadjiev}(2018)}]{Doneva:2017bvd}%
  \BibitemOpen
  \bibfield  {author} {\bibinfo {author} {\bibfnamefont {D.~D.}\ \bibnamefont
  {Doneva}}\ and\ \bibinfo {author} {\bibfnamefont {S.~S.}\ \bibnamefont
  {Yazadjiev}},\ }\href {\doibase 10.1103/PhysRevLett.120.131103} {\bibfield
  {journal} {\bibinfo  {journal} {Phys. Rev. Lett.}\ }\textbf {\bibinfo
  {volume} {120}},\ \bibinfo {pages} {131103} (\bibinfo {year} {2018})},\
  \Eprint {http://arxiv.org/abs/1711.01187} {arXiv:1711.01187 [gr-qc]}
  \BibitemShut {NoStop}%
\bibitem [{\citenamefont {Sotiriou}\ and\ \citenamefont
  {Zhou}(2014{\natexlab{b}})}]{Sotiriou:2014pfa}%
  \BibitemOpen
  \bibfield  {author} {\bibinfo {author} {\bibfnamefont {T.~P.}\ \bibnamefont
  {Sotiriou}}\ and\ \bibinfo {author} {\bibfnamefont {S.-Y.}\ \bibnamefont
  {Zhou}},\ }\href {\doibase 10.1103/PhysRevD.90.124063} {\bibfield  {journal}
  {\bibinfo  {journal} {Phys. Rev. D}\ }\textbf {\bibinfo {volume} {90}},\
  \bibinfo {pages} {124063} (\bibinfo {year} {2014}{\natexlab{b}})},\ \Eprint
  {http://arxiv.org/abs/1408.1698} {arXiv:1408.1698 [gr-qc]} \BibitemShut
  {NoStop}%
\bibitem [{\citenamefont {Yunes}\ and\ \citenamefont
  {Stein}(2011)}]{Yunes:2011we}%
  \BibitemOpen
  \bibfield  {author} {\bibinfo {author} {\bibfnamefont {N.}~\bibnamefont
  {Yunes}}\ and\ \bibinfo {author} {\bibfnamefont {L.~C.}\ \bibnamefont
  {Stein}},\ }\href {\doibase 10.1103/PhysRevD.83.104002} {\bibfield  {journal}
  {\bibinfo  {journal} {Phys. Rev. D}\ }\textbf {\bibinfo {volume} {83}},\
  \bibinfo {pages} {104002} (\bibinfo {year} {2011})},\ \Eprint
  {http://arxiv.org/abs/1101.2921} {arXiv:1101.2921 [gr-qc]} \BibitemShut
  {NoStop}%
\bibitem [{\citenamefont {Maselli}\ \emph {et~al.}(2015)\citenamefont
  {Maselli}, \citenamefont {Pani}, \citenamefont {Gualtieri},\ and\
  \citenamefont {Ferrari}}]{Maselli:2015tta}%
  \BibitemOpen
  \bibfield  {author} {\bibinfo {author} {\bibfnamefont {A.}~\bibnamefont
  {Maselli}}, \bibinfo {author} {\bibfnamefont {P.}~\bibnamefont {Pani}},
  \bibinfo {author} {\bibfnamefont {L.}~\bibnamefont {Gualtieri}}, \ and\
  \bibinfo {author} {\bibfnamefont {V.}~\bibnamefont {Ferrari}},\ }\href
  {\doibase 10.1103/PhysRevD.92.083014} {\bibfield  {journal} {\bibinfo
  {journal} {Phys. Rev. D}\ }\textbf {\bibinfo {volume} {92}},\ \bibinfo
  {pages} {083014} (\bibinfo {year} {2015})},\ \Eprint
  {http://arxiv.org/abs/1507.00680} {arXiv:1507.00680 [gr-qc]} \BibitemShut
  {NoStop}%
\bibitem [{\citenamefont {Dima}\ \emph {et~al.}(2020)\citenamefont {Dima},
  \citenamefont {Barausse}, \citenamefont {Franchini},\ and\ \citenamefont
  {Sotiriou}}]{Dima:2020yac}%
  \BibitemOpen
  \bibfield  {author} {\bibinfo {author} {\bibfnamefont {A.}~\bibnamefont
  {Dima}}, \bibinfo {author} {\bibfnamefont {E.}~\bibnamefont {Barausse}},
  \bibinfo {author} {\bibfnamefont {N.}~\bibnamefont {Franchini}}, \ and\
  \bibinfo {author} {\bibfnamefont {T.~P.}\ \bibnamefont {Sotiriou}},\ }\href
  {\doibase 10.1103/PhysRevLett.125.231101} {\bibfield  {journal} {\bibinfo
  {journal} {Phys. Rev. Lett.}\ }\textbf {\bibinfo {volume} {125}},\ \bibinfo
  {pages} {231101} (\bibinfo {year} {2020})},\ \Eprint
  {http://arxiv.org/abs/2006.03095} {arXiv:2006.03095 [gr-qc]} \BibitemShut
  {NoStop}%
\bibitem [{\citenamefont {Andreou}\ \emph {et~al.}(2019)\citenamefont
  {Andreou}, \citenamefont {Franchini}, \citenamefont {Ventagli},\ and\
  \citenamefont {Sotiriou}}]{Andreou:2019ikc}%
  \BibitemOpen
  \bibfield  {author} {\bibinfo {author} {\bibfnamefont {N.}~\bibnamefont
  {Andreou}}, \bibinfo {author} {\bibfnamefont {N.}~\bibnamefont {Franchini}},
  \bibinfo {author} {\bibfnamefont {G.}~\bibnamefont {Ventagli}}, \ and\
  \bibinfo {author} {\bibfnamefont {T.~P.}\ \bibnamefont {Sotiriou}},\ }\href
  {\doibase 10.1103/PhysRevD.99.124022} {\bibfield  {journal} {\bibinfo
  {journal} {Phys. Rev. D}\ }\textbf {\bibinfo {volume} {99}},\ \bibinfo
  {pages} {124022} (\bibinfo {year} {2019})},\ \bibinfo {note} {[Erratum:
  Phys.Rev.D 101, 109903 (2020)]},\ \Eprint {http://arxiv.org/abs/1904.06365}
  {arXiv:1904.06365 [gr-qc]} \BibitemShut {NoStop}%
\bibitem [{\citenamefont {Bl\'azquez-Salcedo}\ \emph
  {et~al.}(2018)\citenamefont {Bl\'azquez-Salcedo}, \citenamefont {Doneva},
  \citenamefont {Kunz},\ and\ \citenamefont
  {Yazadjiev}}]{Blazquez-Salcedo:2018jnn}%
  \BibitemOpen
  \bibfield  {author} {\bibinfo {author} {\bibfnamefont {J.~L.}\ \bibnamefont
  {Bl\'azquez-Salcedo}}, \bibinfo {author} {\bibfnamefont {D.~D.}\ \bibnamefont
  {Doneva}}, \bibinfo {author} {\bibfnamefont {J.}~\bibnamefont {Kunz}}, \ and\
  \bibinfo {author} {\bibfnamefont {S.~S.}\ \bibnamefont {Yazadjiev}},\ }\href
  {\doibase 10.1103/PhysRevD.98.084011} {\bibfield  {journal} {\bibinfo
  {journal} {Phys. Rev. D}\ }\textbf {\bibinfo {volume} {98}},\ \bibinfo
  {pages} {084011} (\bibinfo {year} {2018})},\ \Eprint
  {http://arxiv.org/abs/1805.05755} {arXiv:1805.05755 [gr-qc]} \BibitemShut
  {NoStop}%
\bibitem [{\citenamefont {Silva}\ \emph {et~al.}(2019)\citenamefont {Silva},
  \citenamefont {Macedo}, \citenamefont {Sotiriou}, \citenamefont {Gualtieri},
  \citenamefont {Sakstein},\ and\ \citenamefont {Berti}}]{Silva:2018qhn}%
  \BibitemOpen
  \bibfield  {author} {\bibinfo {author} {\bibfnamefont {H.~O.}\ \bibnamefont
  {Silva}}, \bibinfo {author} {\bibfnamefont {C.~F.~B.}\ \bibnamefont
  {Macedo}}, \bibinfo {author} {\bibfnamefont {T.~P.}\ \bibnamefont
  {Sotiriou}}, \bibinfo {author} {\bibfnamefont {L.}~\bibnamefont {Gualtieri}},
  \bibinfo {author} {\bibfnamefont {J.}~\bibnamefont {Sakstein}}, \ and\
  \bibinfo {author} {\bibfnamefont {E.}~\bibnamefont {Berti}},\ }\href
  {\doibase 10.1103/PhysRevD.99.064011} {\bibfield  {journal} {\bibinfo
  {journal} {Phys. Rev. D}\ }\textbf {\bibinfo {volume} {99}},\ \bibinfo
  {pages} {064011} (\bibinfo {year} {2019})},\ \Eprint
  {http://arxiv.org/abs/1812.05590} {arXiv:1812.05590 [gr-qc]} \BibitemShut
  {NoStop}%
\bibitem [{\citenamefont {Macedo}\ \emph {et~al.}(2019)\citenamefont {Macedo},
  \citenamefont {Sakstein}, \citenamefont {Berti}, \citenamefont {Gualtieri},
  \citenamefont {Silva},\ and\ \citenamefont {Sotiriou}}]{Macedo:2019sem}%
  \BibitemOpen
  \bibfield  {author} {\bibinfo {author} {\bibfnamefont {C.~F.~B.}\
  \bibnamefont {Macedo}}, \bibinfo {author} {\bibfnamefont {J.}~\bibnamefont
  {Sakstein}}, \bibinfo {author} {\bibfnamefont {E.}~\bibnamefont {Berti}},
  \bibinfo {author} {\bibfnamefont {L.}~\bibnamefont {Gualtieri}}, \bibinfo
  {author} {\bibfnamefont {H.~O.}\ \bibnamefont {Silva}}, \ and\ \bibinfo
  {author} {\bibfnamefont {T.~P.}\ \bibnamefont {Sotiriou}},\ }\href {\doibase
  10.1103/PhysRevD.99.104041} {\bibfield  {journal} {\bibinfo  {journal} {Phys.
  Rev. D}\ }\textbf {\bibinfo {volume} {99}},\ \bibinfo {pages} {104041}
  (\bibinfo {year} {2019})},\ \Eprint {http://arxiv.org/abs/1903.06784}
  {arXiv:1903.06784 [gr-qc]} \BibitemShut {NoStop}%
\bibitem [{\citenamefont {Antoniou}\ \emph
  {et~al.}(2021{\natexlab{a}})\citenamefont {Antoniou}, \citenamefont
  {Leh\'ebel}, \citenamefont {Ventagli},\ and\ \citenamefont
  {Sotiriou}}]{Antoniou:2021zoy}%
  \BibitemOpen
  \bibfield  {author} {\bibinfo {author} {\bibfnamefont {G.}~\bibnamefont
  {Antoniou}}, \bibinfo {author} {\bibfnamefont {A.}~\bibnamefont {Leh\'ebel}},
  \bibinfo {author} {\bibfnamefont {G.}~\bibnamefont {Ventagli}}, \ and\
  \bibinfo {author} {\bibfnamefont {T.~P.}\ \bibnamefont {Sotiriou}},\ }\href
  {\doibase 10.1103/PhysRevD.104.044002} {\bibfield  {journal} {\bibinfo
  {journal} {Phys. Rev. D}\ }\textbf {\bibinfo {volume} {104}},\ \bibinfo
  {pages} {044002} (\bibinfo {year} {2021}{\natexlab{a}})},\ \Eprint
  {http://arxiv.org/abs/2105.04479} {arXiv:2105.04479 [gr-qc]} \BibitemShut
  {NoStop}%
\bibitem [{\citenamefont {Antoniou}\ \emph {et~al.}(2022)\citenamefont
  {Antoniou}, \citenamefont {Macedo}, \citenamefont {McManus},\ and\
  \citenamefont {Sotiriou}}]{Antoniou:2022agj}%
  \BibitemOpen
  \bibfield  {author} {\bibinfo {author} {\bibfnamefont {G.}~\bibnamefont
  {Antoniou}}, \bibinfo {author} {\bibfnamefont {C.~F.~B.}\ \bibnamefont
  {Macedo}}, \bibinfo {author} {\bibfnamefont {R.}~\bibnamefont {McManus}}, \
  and\ \bibinfo {author} {\bibfnamefont {T.~P.}\ \bibnamefont {Sotiriou}},\
  }\href {\doibase 10.1103/PhysRevD.106.024029} {\bibfield  {journal} {\bibinfo
   {journal} {Phys. Rev. D}\ }\textbf {\bibinfo {volume} {106}},\ \bibinfo
  {pages} {024029} (\bibinfo {year} {2022})},\ \Eprint
  {http://arxiv.org/abs/2204.01684} {arXiv:2204.01684 [gr-qc]} \BibitemShut
  {NoStop}%
\bibitem [{\citenamefont {Ayzenberg}\ and\ \citenamefont
  {Yunes}(2014)}]{Ayzenberg:2014aka}%
  \BibitemOpen
  \bibfield  {author} {\bibinfo {author} {\bibfnamefont {D.}~\bibnamefont
  {Ayzenberg}}\ and\ \bibinfo {author} {\bibfnamefont {N.}~\bibnamefont
  {Yunes}},\ }\href {\doibase 10.1103/PhysRevD.90.044066} {\bibfield  {journal}
  {\bibinfo  {journal} {Phys. Rev. D}\ }\textbf {\bibinfo {volume} {90}},\
  \bibinfo {pages} {044066} (\bibinfo {year} {2014})},\ \bibinfo {note}
  {[Erratum: Phys.Rev.D 91, 069905 (2015)]},\ \Eprint
  {http://arxiv.org/abs/1405.2133} {arXiv:1405.2133 [gr-qc]} \BibitemShut
  {NoStop}%
\bibitem [{\citenamefont {Herdeiro}\ \emph {et~al.}(2021)\citenamefont
  {Herdeiro}, \citenamefont {Radu}, \citenamefont {Silva}, \citenamefont
  {Sotiriou},\ and\ \citenamefont {Yunes}}]{Herdeiro:2020wei}%
  \BibitemOpen
  \bibfield  {author} {\bibinfo {author} {\bibfnamefont {C.~A.~R.}\
  \bibnamefont {Herdeiro}}, \bibinfo {author} {\bibfnamefont {E.}~\bibnamefont
  {Radu}}, \bibinfo {author} {\bibfnamefont {H.~O.}\ \bibnamefont {Silva}},
  \bibinfo {author} {\bibfnamefont {T.~P.}\ \bibnamefont {Sotiriou}}, \ and\
  \bibinfo {author} {\bibfnamefont {N.}~\bibnamefont {Yunes}},\ }\href
  {\doibase 10.1103/PhysRevLett.126.011103} {\bibfield  {journal} {\bibinfo
  {journal} {Phys. Rev. Lett.}\ }\textbf {\bibinfo {volume} {126}},\ \bibinfo
  {pages} {011103} (\bibinfo {year} {2021})},\ \Eprint
  {http://arxiv.org/abs/2009.03904} {arXiv:2009.03904 [gr-qc]} \BibitemShut
  {NoStop}%
\bibitem [{\citenamefont {Kanti}\ \emph {et~al.}(1998)\citenamefont {Kanti},
  \citenamefont {Mavromatos}, \citenamefont {Rizos}, \citenamefont {Tamvakis},\
  and\ \citenamefont {Winstanley}}]{Kanti:1997br}%
  \BibitemOpen
  \bibfield  {author} {\bibinfo {author} {\bibfnamefont {P.}~\bibnamefont
  {Kanti}}, \bibinfo {author} {\bibfnamefont {N.~E.}\ \bibnamefont
  {Mavromatos}}, \bibinfo {author} {\bibfnamefont {J.}~\bibnamefont {Rizos}},
  \bibinfo {author} {\bibfnamefont {K.}~\bibnamefont {Tamvakis}}, \ and\
  \bibinfo {author} {\bibfnamefont {E.}~\bibnamefont {Winstanley}},\ }\href
  {\doibase 10.1103/PhysRevD.57.6255} {\bibfield  {journal} {\bibinfo
  {journal} {Phys. Rev. D}\ }\textbf {\bibinfo {volume} {57}},\ \bibinfo
  {pages} {6255} (\bibinfo {year} {1998})},\ \Eprint
  {http://arxiv.org/abs/hep-th/9703192} {arXiv:hep-th/9703192} \BibitemShut
  {NoStop}%
\bibitem [{\citenamefont {Macedo}(2020)}]{Macedo:2020tbm}%
  \BibitemOpen
  \bibfield  {author} {\bibinfo {author} {\bibfnamefont {C.~F.~B.}\
  \bibnamefont {Macedo}},\ }\href {\doibase 10.1142/S0218271820410060}
  {\bibfield  {journal} {\bibinfo  {journal} {Int. J. Mod. Phys. D}\ }\textbf
  {\bibinfo {volume} {29}},\ \bibinfo {pages} {2041006} (\bibinfo {year}
  {2020})},\ \Eprint {http://arxiv.org/abs/2002.12719} {arXiv:2002.12719
  [gr-qc]} \BibitemShut {NoStop}%
\bibitem [{\citenamefont {Bl\'azquez-Salcedo}\ \emph
  {et~al.}(2020{\natexlab{a}})\citenamefont {Bl\'azquez-Salcedo}, \citenamefont
  {Doneva}, \citenamefont {Kahlen}, \citenamefont {Kunz}, \citenamefont
  {Nedkova},\ and\ \citenamefont {Yazadjiev}}]{Blazquez-Salcedo:2020rhf}%
  \BibitemOpen
  \bibfield  {author} {\bibinfo {author} {\bibfnamefont {J.~L.}\ \bibnamefont
  {Bl\'azquez-Salcedo}}, \bibinfo {author} {\bibfnamefont {D.~D.}\ \bibnamefont
  {Doneva}}, \bibinfo {author} {\bibfnamefont {S.}~\bibnamefont {Kahlen}},
  \bibinfo {author} {\bibfnamefont {J.}~\bibnamefont {Kunz}}, \bibinfo {author}
  {\bibfnamefont {P.}~\bibnamefont {Nedkova}}, \ and\ \bibinfo {author}
  {\bibfnamefont {S.~S.}\ \bibnamefont {Yazadjiev}},\ }\href {\doibase
  10.1103/PhysRevD.101.104006} {\bibfield  {journal} {\bibinfo  {journal}
  {Phys. Rev. D}\ }\textbf {\bibinfo {volume} {101}},\ \bibinfo {pages}
  {104006} (\bibinfo {year} {2020}{\natexlab{a}})},\ \Eprint
  {http://arxiv.org/abs/2003.02862} {arXiv:2003.02862 [gr-qc]} \BibitemShut
  {NoStop}%
\bibitem [{\citenamefont {Minamitsuji}\ \emph {et~al.}(2024)\citenamefont
  {Minamitsuji}, \citenamefont {Mukohyama},\ and\ \citenamefont
  {Tsujikawa}}]{Minamitsuji:2024twp}%
  \BibitemOpen
  \bibfield  {author} {\bibinfo {author} {\bibfnamefont {M.}~\bibnamefont
  {Minamitsuji}}, \bibinfo {author} {\bibfnamefont {S.}~\bibnamefont
  {Mukohyama}}, \ and\ \bibinfo {author} {\bibfnamefont {S.}~\bibnamefont
  {Tsujikawa}},\ }\href@noop {} {\  (\bibinfo {year} {2024})},\ \Eprint
  {http://arxiv.org/abs/2403.10048} {arXiv:2403.10048 [gr-qc]} \BibitemShut
  {NoStop}%
\bibitem [{\citenamefont {Wong}\ \emph {et~al.}(2022)\citenamefont {Wong},
  \citenamefont {Herdeiro},\ and\ \citenamefont {Radu}}]{Wong:2022wni}%
  \BibitemOpen
  \bibfield  {author} {\bibinfo {author} {\bibfnamefont {L.~K.}\ \bibnamefont
  {Wong}}, \bibinfo {author} {\bibfnamefont {C.~A.~R.}\ \bibnamefont
  {Herdeiro}}, \ and\ \bibinfo {author} {\bibfnamefont {E.}~\bibnamefont
  {Radu}},\ }\href {\doibase 10.1103/PhysRevD.106.024008} {\bibfield  {journal}
  {\bibinfo  {journal} {Phys. Rev. D}\ }\textbf {\bibinfo {volume} {106}},\
  \bibinfo {pages} {024008} (\bibinfo {year} {2022})},\ \Eprint
  {http://arxiv.org/abs/2204.09038} {arXiv:2204.09038 [gr-qc]} \BibitemShut
  {NoStop}%
\bibitem [{\citenamefont {Ripley}\ and\ \citenamefont
  {Pretorius}(2020)}]{Ripley:2019aqj}%
  \BibitemOpen
  \bibfield  {author} {\bibinfo {author} {\bibfnamefont {J.~L.}\ \bibnamefont
  {Ripley}}\ and\ \bibinfo {author} {\bibfnamefont {F.}~\bibnamefont
  {Pretorius}},\ }\href {\doibase 10.1103/PhysRevD.101.044015} {\bibfield
  {journal} {\bibinfo  {journal} {Phys. Rev. D}\ }\textbf {\bibinfo {volume}
  {101}},\ \bibinfo {pages} {044015} (\bibinfo {year} {2020})},\ \Eprint
  {http://arxiv.org/abs/1911.11027} {arXiv:1911.11027 [gr-qc]} \BibitemShut
  {NoStop}%
\bibitem [{\citenamefont {East}\ and\ \citenamefont
  {Ripley}(2021{\natexlab{a}})}]{East:2020hgw}%
  \BibitemOpen
  \bibfield  {author} {\bibinfo {author} {\bibfnamefont {W.~E.}\ \bibnamefont
  {East}}\ and\ \bibinfo {author} {\bibfnamefont {J.~L.}\ \bibnamefont
  {Ripley}},\ }\href {\doibase 10.1103/PhysRevD.103.044040} {\bibfield
  {journal} {\bibinfo  {journal} {Phys. Rev. D}\ }\textbf {\bibinfo {volume}
  {103}},\ \bibinfo {pages} {044040} (\bibinfo {year} {2021}{\natexlab{a}})},\
  \Eprint {http://arxiv.org/abs/2011.03547} {arXiv:2011.03547 [gr-qc]}
  \BibitemShut {NoStop}%
\bibitem [{\citenamefont {East}\ and\ \citenamefont
  {Ripley}(2021{\natexlab{b}})}]{East:2021bqk}%
  \BibitemOpen
  \bibfield  {author} {\bibinfo {author} {\bibfnamefont {W.~E.}\ \bibnamefont
  {East}}\ and\ \bibinfo {author} {\bibfnamefont {J.~L.}\ \bibnamefont
  {Ripley}},\ }\href {\doibase 10.1103/PhysRevLett.127.101102} {\bibfield
  {journal} {\bibinfo  {journal} {Phys. Rev. Lett.}\ }\textbf {\bibinfo
  {volume} {127}},\ \bibinfo {pages} {101102} (\bibinfo {year}
  {2021}{\natexlab{b}})},\ \Eprint {http://arxiv.org/abs/2105.08571}
  {arXiv:2105.08571 [gr-qc]} \BibitemShut {NoStop}%
\bibitem [{\citenamefont {Doneva}\ \emph {et~al.}(2024)\citenamefont {Doneva},
  \citenamefont {Ramazano\u{g}lu}, \citenamefont {Silva}, \citenamefont
  {Sotiriou},\ and\ \citenamefont {Yazadjiev}}]{Doneva:2022ewd}%
  \BibitemOpen
  \bibfield  {author} {\bibinfo {author} {\bibfnamefont {D.~D.}\ \bibnamefont
  {Doneva}}, \bibinfo {author} {\bibfnamefont {F.~M.}\ \bibnamefont
  {Ramazano\u{g}lu}}, \bibinfo {author} {\bibfnamefont {H.~O.}\ \bibnamefont
  {Silva}}, \bibinfo {author} {\bibfnamefont {T.~P.}\ \bibnamefont {Sotiriou}},
  \ and\ \bibinfo {author} {\bibfnamefont {S.~S.}\ \bibnamefont {Yazadjiev}},\
  }\href {\doibase 10.1103/RevModPhys.96.015004} {\bibfield  {journal}
  {\bibinfo  {journal} {Rev. Mod. Phys.}\ }\textbf {\bibinfo {volume} {96}},\
  \bibinfo {pages} {015004} (\bibinfo {year} {2024})},\ \Eprint
  {http://arxiv.org/abs/2211.01766} {arXiv:2211.01766 [gr-qc]} \BibitemShut
  {NoStop}%
\bibitem [{\citenamefont {Ventagli}\ \emph {et~al.}(2021)\citenamefont
  {Ventagli}, \citenamefont {Antoniou}, \citenamefont {Leh\'ebel},\ and\
  \citenamefont {Sotiriou}}]{Ventagli:2021ubn}%
  \BibitemOpen
  \bibfield  {author} {\bibinfo {author} {\bibfnamefont {G.}~\bibnamefont
  {Ventagli}}, \bibinfo {author} {\bibfnamefont {G.}~\bibnamefont {Antoniou}},
  \bibinfo {author} {\bibfnamefont {A.}~\bibnamefont {Leh\'ebel}}, \ and\
  \bibinfo {author} {\bibfnamefont {T.~P.}\ \bibnamefont {Sotiriou}},\ }\href
  {\doibase 10.1103/PhysRevD.104.124078} {\bibfield  {journal} {\bibinfo
  {journal} {Phys. Rev. D}\ }\textbf {\bibinfo {volume} {104}},\ \bibinfo
  {pages} {124078} (\bibinfo {year} {2021})},\ \Eprint
  {http://arxiv.org/abs/2111.03644} {arXiv:2111.03644 [gr-qc]} \BibitemShut
  {NoStop}%
\bibitem [{\citenamefont {Antoniou}\ \emph
  {et~al.}(2021{\natexlab{b}})\citenamefont {Antoniou}, \citenamefont
  {Bordin},\ and\ \citenamefont {Sotiriou}}]{Antoniou:2020nax}%
  \BibitemOpen
  \bibfield  {author} {\bibinfo {author} {\bibfnamefont {G.}~\bibnamefont
  {Antoniou}}, \bibinfo {author} {\bibfnamefont {L.}~\bibnamefont {Bordin}}, \
  and\ \bibinfo {author} {\bibfnamefont {T.~P.}\ \bibnamefont {Sotiriou}},\
  }\href {\doibase 10.1103/PhysRevD.103.024012} {\bibfield  {journal} {\bibinfo
   {journal} {Phys. Rev. D}\ }\textbf {\bibinfo {volume} {103}},\ \bibinfo
  {pages} {024012} (\bibinfo {year} {2021}{\natexlab{b}})},\ \Eprint
  {http://arxiv.org/abs/2004.14985} {arXiv:2004.14985 [gr-qc]} \BibitemShut
  {NoStop}%
\bibitem [{\citenamefont {Thaalba}\ \emph
  {et~al.}(2023{\natexlab{a}})\citenamefont {Thaalba}, \citenamefont {Bezares},
  \citenamefont {Franchini},\ and\ \citenamefont {Sotiriou}}]{Thaalba:2023fmq}%
  \BibitemOpen
  \bibfield  {author} {\bibinfo {author} {\bibfnamefont {F.}~\bibnamefont
  {Thaalba}}, \bibinfo {author} {\bibfnamefont {M.}~\bibnamefont {Bezares}},
  \bibinfo {author} {\bibfnamefont {N.}~\bibnamefont {Franchini}}, \ and\
  \bibinfo {author} {\bibfnamefont {T.~P.}\ \bibnamefont {Sotiriou}},\
  }\href@noop {} {\  (\bibinfo {year} {2023}{\natexlab{a}})},\ \Eprint
  {http://arxiv.org/abs/2306.01695} {arXiv:2306.01695 [gr-qc]} \BibitemShut
  {NoStop}%
\bibitem [{\citenamefont {Bl\'azquez-Salcedo}\ \emph
  {et~al.}(2017)\citenamefont {Bl\'azquez-Salcedo}, \citenamefont {Khoo},\ and\
  \citenamefont {Kunz}}]{Blazquez-Salcedo:2017txk}%
  \BibitemOpen
  \bibfield  {author} {\bibinfo {author} {\bibfnamefont {J.~L.}\ \bibnamefont
  {Bl\'azquez-Salcedo}}, \bibinfo {author} {\bibfnamefont {F.~S.}\ \bibnamefont
  {Khoo}}, \ and\ \bibinfo {author} {\bibfnamefont {J.}~\bibnamefont {Kunz}},\
  }\href {\doibase 10.1103/PhysRevD.96.064008} {\bibfield  {journal} {\bibinfo
  {journal} {Phys. Rev. D}\ }\textbf {\bibinfo {volume} {96}},\ \bibinfo
  {pages} {064008} (\bibinfo {year} {2017})},\ \Eprint
  {http://arxiv.org/abs/1706.03262} {arXiv:1706.03262 [gr-qc]} \BibitemShut
  {NoStop}%
\bibitem [{\citenamefont {Bl\'azquez-Salcedo}\ \emph
  {et~al.}(2016)\citenamefont {Bl\'azquez-Salcedo}, \citenamefont {Macedo},
  \citenamefont {Cardoso}, \citenamefont {Ferrari}, \citenamefont {Gualtieri},
  \citenamefont {Khoo}, \citenamefont {Kunz},\ and\ \citenamefont
  {Pani}}]{Blazquez-Salcedo:2016enn}%
  \BibitemOpen
  \bibfield  {author} {\bibinfo {author} {\bibfnamefont {J.~L.}\ \bibnamefont
  {Bl\'azquez-Salcedo}}, \bibinfo {author} {\bibfnamefont {C.~F.~B.}\
  \bibnamefont {Macedo}}, \bibinfo {author} {\bibfnamefont {V.}~\bibnamefont
  {Cardoso}}, \bibinfo {author} {\bibfnamefont {V.}~\bibnamefont {Ferrari}},
  \bibinfo {author} {\bibfnamefont {L.}~\bibnamefont {Gualtieri}}, \bibinfo
  {author} {\bibfnamefont {F.~S.}\ \bibnamefont {Khoo}}, \bibinfo {author}
  {\bibfnamefont {J.}~\bibnamefont {Kunz}}, \ and\ \bibinfo {author}
  {\bibfnamefont {P.}~\bibnamefont {Pani}},\ }\href {\doibase
  10.1103/PhysRevD.94.104024} {\bibfield  {journal} {\bibinfo  {journal} {Phys.
  Rev. D}\ }\textbf {\bibinfo {volume} {94}},\ \bibinfo {pages} {104024}
  (\bibinfo {year} {2016})},\ \Eprint {http://arxiv.org/abs/1609.01286}
  {arXiv:1609.01286 [gr-qc]} \BibitemShut {NoStop}%
\bibitem [{\citenamefont {Pierini}\ and\ \citenamefont
  {Gualtieri}(2022)}]{Pierini:2022eim}%
  \BibitemOpen
  \bibfield  {author} {\bibinfo {author} {\bibfnamefont {L.}~\bibnamefont
  {Pierini}}\ and\ \bibinfo {author} {\bibfnamefont {L.}~\bibnamefont
  {Gualtieri}},\ }\href {\doibase 10.1103/PhysRevD.106.104009} {\bibfield
  {journal} {\bibinfo  {journal} {Phys. Rev. D}\ }\textbf {\bibinfo {volume}
  {106}},\ \bibinfo {pages} {104009} (\bibinfo {year} {2022})},\ \Eprint
  {http://arxiv.org/abs/2207.11267} {arXiv:2207.11267 [gr-qc]} \BibitemShut
  {NoStop}%
\bibitem [{\citenamefont {Bl\'azquez-Salcedo}\ \emph
  {et~al.}(2020{\natexlab{b}})\citenamefont {Bl\'azquez-Salcedo}, \citenamefont
  {Doneva}, \citenamefont {Kahlen}, \citenamefont {Kunz}, \citenamefont
  {Nedkova},\ and\ \citenamefont {Yazadjiev}}]{Blazquez-Salcedo:2020caw}%
  \BibitemOpen
  \bibfield  {author} {\bibinfo {author} {\bibfnamefont {J.~L.}\ \bibnamefont
  {Bl\'azquez-Salcedo}}, \bibinfo {author} {\bibfnamefont {D.~D.}\ \bibnamefont
  {Doneva}}, \bibinfo {author} {\bibfnamefont {S.}~\bibnamefont {Kahlen}},
  \bibinfo {author} {\bibfnamefont {J.}~\bibnamefont {Kunz}}, \bibinfo {author}
  {\bibfnamefont {P.}~\bibnamefont {Nedkova}}, \ and\ \bibinfo {author}
  {\bibfnamefont {S.~S.}\ \bibnamefont {Yazadjiev}},\ }\href {\doibase
  10.1103/PhysRevD.102.024086} {\bibfield  {journal} {\bibinfo  {journal}
  {Phys. Rev. D}\ }\textbf {\bibinfo {volume} {102}},\ \bibinfo {pages}
  {024086} (\bibinfo {year} {2020}{\natexlab{b}})},\ \Eprint
  {http://arxiv.org/abs/2006.06006} {arXiv:2006.06006 [gr-qc]} \BibitemShut
  {NoStop}%
\bibitem [{\citenamefont {Antoniou}\ \emph
  {et~al.}(2018{\natexlab{b}})\citenamefont {Antoniou}, \citenamefont
  {Bakopoulos},\ and\ \citenamefont {Kanti}}]{Antoniou:2017hxj}%
  \BibitemOpen
  \bibfield  {author} {\bibinfo {author} {\bibfnamefont {G.}~\bibnamefont
  {Antoniou}}, \bibinfo {author} {\bibfnamefont {A.}~\bibnamefont
  {Bakopoulos}}, \ and\ \bibinfo {author} {\bibfnamefont {P.}~\bibnamefont
  {Kanti}},\ }\href {\doibase 10.1103/PhysRevD.97.084037} {\bibfield  {journal}
  {\bibinfo  {journal} {Phys. Rev. D}\ }\textbf {\bibinfo {volume} {97}},\
  \bibinfo {pages} {084037} (\bibinfo {year} {2018}{\natexlab{b}})},\ \Eprint
  {http://arxiv.org/abs/1711.07431} {arXiv:1711.07431 [hep-th]} \BibitemShut
  {NoStop}%
\bibitem [{\citenamefont {Bakopoulos}\ \emph {et~al.}(2019)\citenamefont
  {Bakopoulos}, \citenamefont {Antoniou},\ and\ \citenamefont
  {Kanti}}]{Bakopoulos:2018nui}%
  \BibitemOpen
  \bibfield  {author} {\bibinfo {author} {\bibfnamefont {A.}~\bibnamefont
  {Bakopoulos}}, \bibinfo {author} {\bibfnamefont {G.}~\bibnamefont
  {Antoniou}}, \ and\ \bibinfo {author} {\bibfnamefont {P.}~\bibnamefont
  {Kanti}},\ }\href {\doibase 10.1103/PhysRevD.99.064003} {\bibfield  {journal}
  {\bibinfo  {journal} {Phys. Rev. D}\ }\textbf {\bibinfo {volume} {99}},\
  \bibinfo {pages} {064003} (\bibinfo {year} {2019})},\ \Eprint
  {http://arxiv.org/abs/1812.06941} {arXiv:1812.06941 [hep-th]} \BibitemShut
  {NoStop}%
\bibitem [{\citenamefont {Ramazano\u{g}lu}\ and\ \citenamefont
  {Pretorius}(2016)}]{Ramazanoglu:2016kul}%
  \BibitemOpen
  \bibfield  {author} {\bibinfo {author} {\bibfnamefont {F.~M.}\ \bibnamefont
  {Ramazano\u{g}lu}}\ and\ \bibinfo {author} {\bibfnamefont {F.}~\bibnamefont
  {Pretorius}},\ }\href {\doibase 10.1103/PhysRevD.93.064005} {\bibfield
  {journal} {\bibinfo  {journal} {Phys. Rev. D}\ }\textbf {\bibinfo {volume}
  {93}},\ \bibinfo {pages} {064005} (\bibinfo {year} {2016})},\ \Eprint
  {http://arxiv.org/abs/1601.07475} {arXiv:1601.07475 [gr-qc]} \BibitemShut
  {NoStop}%
\bibitem [{\citenamefont {Herdeiro}\ \emph {et~al.}(2018)\citenamefont
  {Herdeiro}, \citenamefont {Radu}, \citenamefont {Sanchis-Gual},\ and\
  \citenamefont {Font}}]{Herdeiro:2018wub}%
  \BibitemOpen
  \bibfield  {author} {\bibinfo {author} {\bibfnamefont {C.~A.~R.}\
  \bibnamefont {Herdeiro}}, \bibinfo {author} {\bibfnamefont {E.}~\bibnamefont
  {Radu}}, \bibinfo {author} {\bibfnamefont {N.}~\bibnamefont {Sanchis-Gual}},
  \ and\ \bibinfo {author} {\bibfnamefont {J.~A.}\ \bibnamefont {Font}},\
  }\href {\doibase 10.1103/PhysRevLett.121.101102} {\bibfield  {journal}
  {\bibinfo  {journal} {Phys. Rev. Lett.}\ }\textbf {\bibinfo {volume} {121}},\
  \bibinfo {pages} {101102} (\bibinfo {year} {2018})},\ \Eprint
  {http://arxiv.org/abs/1806.05190} {arXiv:1806.05190 [gr-qc]} \BibitemShut
  {NoStop}%
\bibitem [{\citenamefont {Ramazano\u{g}lu}(2017)}]{Ramazanoglu:2017xbl}%
  \BibitemOpen
  \bibfield  {author} {\bibinfo {author} {\bibfnamefont {F.~M.}\ \bibnamefont
  {Ramazano\u{g}lu}},\ }\href {\doibase 10.1103/PhysRevD.96.064009} {\bibfield
  {journal} {\bibinfo  {journal} {Phys. Rev. D}\ }\textbf {\bibinfo {volume}
  {96}},\ \bibinfo {pages} {064009} (\bibinfo {year} {2017})},\ \Eprint
  {http://arxiv.org/abs/1706.01056} {arXiv:1706.01056 [gr-qc]} \BibitemShut
  {NoStop}%
\bibitem [{\citenamefont {Ramazano\u{g}lu}(2018)}]{Ramazanoglu:2018hwk}%
  \BibitemOpen
  \bibfield  {author} {\bibinfo {author} {\bibfnamefont {F.~M.}\ \bibnamefont
  {Ramazano\u{g}lu}},\ }\href {\doibase 10.1103/PhysRevD.98.044011} {\bibfield
  {journal} {\bibinfo  {journal} {Phys. Rev. D}\ }\textbf {\bibinfo {volume}
  {98}},\ \bibinfo {pages} {044011} (\bibinfo {year} {2018})},\ \bibinfo {note}
  {[Erratum: Phys.Rev.D 100, 029903 (2019)]},\ \Eprint
  {http://arxiv.org/abs/1804.00594} {arXiv:1804.00594 [gr-qc]} \BibitemShut
  {NoStop}%
\bibitem [{\citenamefont {Berti}\ \emph {et~al.}(2021)\citenamefont {Berti},
  \citenamefont {Collodel}, \citenamefont {Kleihaus},\ and\ \citenamefont
  {Kunz}}]{Berti:2020kgk}%
  \BibitemOpen
  \bibfield  {author} {\bibinfo {author} {\bibfnamefont {E.}~\bibnamefont
  {Berti}}, \bibinfo {author} {\bibfnamefont {L.~G.}\ \bibnamefont {Collodel}},
  \bibinfo {author} {\bibfnamefont {B.}~\bibnamefont {Kleihaus}}, \ and\
  \bibinfo {author} {\bibfnamefont {J.}~\bibnamefont {Kunz}},\ }\href {\doibase
  10.1103/PhysRevLett.126.011104} {\bibfield  {journal} {\bibinfo  {journal}
  {Phys. Rev. Lett.}\ }\textbf {\bibinfo {volume} {126}},\ \bibinfo {pages}
  {011104} (\bibinfo {year} {2021})},\ \Eprint
  {http://arxiv.org/abs/2009.03905} {arXiv:2009.03905 [gr-qc]} \BibitemShut
  {NoStop}%
\bibitem [{\citenamefont {Thaalba}\ \emph
  {et~al.}(2023{\natexlab{b}})\citenamefont {Thaalba}, \citenamefont
  {Antoniou},\ and\ \citenamefont {Sotiriou}}]{Thaalba:2022bnt}%
  \BibitemOpen
  \bibfield  {author} {\bibinfo {author} {\bibfnamefont {F.}~\bibnamefont
  {Thaalba}}, \bibinfo {author} {\bibfnamefont {G.}~\bibnamefont {Antoniou}}, \
  and\ \bibinfo {author} {\bibfnamefont {T.~P.}\ \bibnamefont {Sotiriou}},\
  }\href {\doibase 10.1088/1361-6382/acdd42} {\bibfield  {journal} {\bibinfo
  {journal} {Class. Quant. Grav.}\ }\textbf {\bibinfo {volume} {40}},\ \bibinfo
  {pages} {155002} (\bibinfo {year} {2023}{\natexlab{b}})},\ \Eprint
  {http://arxiv.org/abs/2211.05099} {arXiv:2211.05099 [gr-qc]} \BibitemShut
  {NoStop}%
\bibitem [{\citenamefont {Regge}\ and\ \citenamefont
  {Wheeler}(1957)}]{Regge:1957td}%
  \BibitemOpen
  \bibfield  {author} {\bibinfo {author} {\bibfnamefont {T.}~\bibnamefont
  {Regge}}\ and\ \bibinfo {author} {\bibfnamefont {J.~A.}\ \bibnamefont
  {Wheeler}},\ }\href {\doibase 10.1103/PhysRev.108.1063} {\bibfield  {journal}
  {\bibinfo  {journal} {Phys. Rev.}\ }\textbf {\bibinfo {volume} {108}},\
  \bibinfo {pages} {1063} (\bibinfo {year} {1957})}\BibitemShut {NoStop}%
\bibitem [{\citenamefont {Zerilli}(1970)}]{Zerilli:1970wzz}%
  \BibitemOpen
  \bibfield  {author} {\bibinfo {author} {\bibfnamefont {F.~J.}\ \bibnamefont
  {Zerilli}},\ }\href {\doibase 10.1103/PhysRevD.2.2141} {\bibfield  {journal}
  {\bibinfo  {journal} {Phys. Rev. D}\ }\textbf {\bibinfo {volume} {2}},\
  \bibinfo {pages} {2141} (\bibinfo {year} {1970})}\BibitemShut {NoStop}%
\bibitem [{\citenamefont {Gundlach}\ \emph {et~al.}(1994)\citenamefont
  {Gundlach}, \citenamefont {Price},\ and\ \citenamefont
  {Pullin}}]{Gundlach:1993tp}%
  \BibitemOpen
  \bibfield  {author} {\bibinfo {author} {\bibfnamefont {C.}~\bibnamefont
  {Gundlach}}, \bibinfo {author} {\bibfnamefont {R.~H.}\ \bibnamefont {Price}},
  \ and\ \bibinfo {author} {\bibfnamefont {J.}~\bibnamefont {Pullin}},\ }\href
  {\doibase 10.1103/PhysRevD.49.883} {\bibfield  {journal} {\bibinfo  {journal}
  {Phys. Rev. D}\ }\textbf {\bibinfo {volume} {49}},\ \bibinfo {pages} {883}
  (\bibinfo {year} {1994})},\ \Eprint {http://arxiv.org/abs/gr-qc/9307009}
  {arXiv:gr-qc/9307009} \BibitemShut {NoStop}%
\bibitem [{\citenamefont {Konoplya}\ and\ \citenamefont
  {Zhidenko}(2011)}]{Konoplya:2011qq}%
  \BibitemOpen
  \bibfield  {author} {\bibinfo {author} {\bibfnamefont {R.~A.}\ \bibnamefont
  {Konoplya}}\ and\ \bibinfo {author} {\bibfnamefont {A.}~\bibnamefont
  {Zhidenko}},\ }\href {\doibase 10.1103/RevModPhys.83.793} {\bibfield
  {journal} {\bibinfo  {journal} {Rev. Mod. Phys.}\ }\textbf {\bibinfo {volume}
  {83}},\ \bibinfo {pages} {793} (\bibinfo {year} {2011})},\ \Eprint
  {http://arxiv.org/abs/1102.4014} {arXiv:1102.4014 [gr-qc]} \BibitemShut
  {NoStop}%
\bibitem [{\citenamefont {Boyce}\ and\ \citenamefont
  {DiPrima}(2004)}]{boyce2004ede}%
  \BibitemOpen
  \bibfield  {author} {\bibinfo {author} {\bibfnamefont {W.}~\bibnamefont
  {Boyce}}\ and\ \bibinfo {author} {\bibfnamefont {R.}~\bibnamefont
  {DiPrima}},\ }\href@noop {} {\emph {\bibinfo {title} {{Elementary
  differential equations and boundary value problems}}}},\ \bibinfo {edition}
  {8th}\ ed.\ (\bibinfo  {publisher} {Wiley New York},\ \bibinfo {year}
  {2004})\BibitemShut {NoStop}%
\bibitem [{\citenamefont {Kimura}(2017)}]{Kimura:2017uor}%
  \BibitemOpen
  \bibfield  {author} {\bibinfo {author} {\bibfnamefont {M.}~\bibnamefont
  {Kimura}},\ }\href {\doibase 10.1088/1361-6382/aa903f} {\bibfield  {journal}
  {\bibinfo  {journal} {Class. Quant. Grav.}\ }\textbf {\bibinfo {volume}
  {34}},\ \bibinfo {pages} {235007} (\bibinfo {year} {2017})},\ \Eprint
  {http://arxiv.org/abs/1706.01447} {arXiv:1706.01447 [gr-qc]} \BibitemShut
  {NoStop}%
\bibitem [{\citenamefont {Kimura}\ and\ \citenamefont
  {Tanaka}(2019)}]{Kimura:2018whv}%
  \BibitemOpen
  \bibfield  {author} {\bibinfo {author} {\bibfnamefont {M.}~\bibnamefont
  {Kimura}}\ and\ \bibinfo {author} {\bibfnamefont {T.}~\bibnamefont
  {Tanaka}},\ }\href {\doibase 10.1088/1361-6382/ab0193} {\bibfield  {journal}
  {\bibinfo  {journal} {Class. Quant. Grav.}\ }\textbf {\bibinfo {volume}
  {36}},\ \bibinfo {pages} {055005} (\bibinfo {year} {2019})},\ \Eprint
  {http://arxiv.org/abs/1809.00795} {arXiv:1809.00795 [gr-qc]} \BibitemShut
  {NoStop}%
\bibitem [{\citenamefont {Bl\'azquez-Salcedo}\ \emph
  {et~al.}(2019)\citenamefont {Bl\'azquez-Salcedo}, \citenamefont {Kahlen},\
  and\ \citenamefont {Kunz}}]{Blazquez-Salcedo:2019nwd}%
  \BibitemOpen
  \bibfield  {author} {\bibinfo {author} {\bibfnamefont {J.~L.}\ \bibnamefont
  {Bl\'azquez-Salcedo}}, \bibinfo {author} {\bibfnamefont {S.}~\bibnamefont
  {Kahlen}}, \ and\ \bibinfo {author} {\bibfnamefont {J.}~\bibnamefont
  {Kunz}},\ }\href {\doibase 10.1140/epjc/s10052-019-7535-4} {\bibfield
  {journal} {\bibinfo  {journal} {Eur. Phys. J. C}\ }\textbf {\bibinfo {volume}
  {79}},\ \bibinfo {pages} {1021} (\bibinfo {year} {2019})},\ \Eprint
  {http://arxiv.org/abs/1911.01943} {arXiv:1911.01943 [gr-qc]} \BibitemShut
  {NoStop}%
\bibitem [{\citenamefont {Wald}(1984)}]{Wald:1984rg}%
  \BibitemOpen
  \bibfield  {author} {\bibinfo {author} {\bibfnamefont {R.~M.}\ \bibnamefont
  {Wald}},\ }\href {\doibase 10.7208/chicago/9780226870373.001.0001} {\emph
  {\bibinfo {title} {{General Relativity}}}}\ (\bibinfo  {publisher} {Chicago
  Univ. Pr.},\ \bibinfo {address} {Chicago, USA},\ \bibinfo {year}
  {1984})\BibitemShut {NoStop}%
\bibitem [{\citenamefont {Cheung}\ \emph {et~al.}(2022)\citenamefont {Cheung},
  \citenamefont {Destounis}, \citenamefont {Macedo}, \citenamefont {Berti},\
  and\ \citenamefont {Cardoso}}]{Cheung:2021bol}%
  \BibitemOpen
  \bibfield  {author} {\bibinfo {author} {\bibfnamefont {M.~H.-Y.}\
  \bibnamefont {Cheung}}, \bibinfo {author} {\bibfnamefont {K.}~\bibnamefont
  {Destounis}}, \bibinfo {author} {\bibfnamefont {R.~P.}\ \bibnamefont
  {Macedo}}, \bibinfo {author} {\bibfnamefont {E.}~\bibnamefont {Berti}}, \
  and\ \bibinfo {author} {\bibfnamefont {V.}~\bibnamefont {Cardoso}},\ }\href
  {\doibase 10.1103/PhysRevLett.128.111103} {\bibfield  {journal} {\bibinfo
  {journal} {Phys. Rev. Lett.}\ }\textbf {\bibinfo {volume} {128}},\ \bibinfo
  {pages} {111103} (\bibinfo {year} {2022})},\ \Eprint
  {http://arxiv.org/abs/2111.05415} {arXiv:2111.05415 [gr-qc]} \BibitemShut
  {NoStop}%
\bibitem [{\citenamefont {Maselli}\ \emph {et~al.}(2024)\citenamefont
  {Maselli}, \citenamefont {Yi}, \citenamefont {Pierini}, \citenamefont
  {Vellucci}, \citenamefont {Reali}, \citenamefont {Gualtieri},\ and\
  \citenamefont {Berti}}]{Maselli:2023khq}%
  \BibitemOpen
  \bibfield  {author} {\bibinfo {author} {\bibfnamefont {A.}~\bibnamefont
  {Maselli}}, \bibinfo {author} {\bibfnamefont {S.}~\bibnamefont {Yi}},
  \bibinfo {author} {\bibfnamefont {L.}~\bibnamefont {Pierini}}, \bibinfo
  {author} {\bibfnamefont {V.}~\bibnamefont {Vellucci}}, \bibinfo {author}
  {\bibfnamefont {L.}~\bibnamefont {Reali}}, \bibinfo {author} {\bibfnamefont
  {L.}~\bibnamefont {Gualtieri}}, \ and\ \bibinfo {author} {\bibfnamefont
  {E.}~\bibnamefont {Berti}},\ }\href {\doibase 10.1103/PhysRevD.109.064060}
  {\bibfield  {journal} {\bibinfo  {journal} {Phys. Rev. D}\ }\textbf {\bibinfo
  {volume} {109}},\ \bibinfo {pages} {064060} (\bibinfo {year} {2024})},\
  \Eprint {http://arxiv.org/abs/2311.14803} {arXiv:2311.14803 [gr-qc]}
  \BibitemShut {NoStop}%
\end{thebibliography}%

\end{document}